\documentclass[a4paper,11pt]{article}
\usepackage{jheppub} 
\usepackage{graphicx}
\usepackage{slashed}

\newcommand{\MET}{\slashed{E}_T}
\newcommand{\mDM}{m_{\rm{DM}}}
\newcommand{\mMed}{M_{\rm{med}}}
\newcommand{\gDM}{g_{\rm{DM}}}
\newcommand{\gq}{g_q}
\newcommand{\ifb}{\rm{fb}^{-1}}

\title{
Characterising dark matter searches at colliders and direct detection experiments: Vector mediators
}

\arxivnumber{ IPPP/14/68, DCPT/14/136, SLAC-PUB-16033}

\author[a]{Oliver Buchmueller,}
\author[b]{Matthew J.~Dolan,}
\author[a]{Sarah A.~Malik}
\author[c,d]{and Christopher McCabe}

\affiliation[a]{High Energy Physics Group, Blackett Laboratory, Imperial College, Prince Consort Road, London, SW7 2AZ, UK\ }
\affiliation[b]{Theory Group, SLAC National Accelerator Laboratory, Menlo Park, California 94025, USA\ }
\affiliation[c]{Institute for Particle Physics Phenomenology, Durham University, South Road, Durham, DH1 3LE, UK\ }
\affiliation[d]{GRAPPA, University of Amsterdam, Science Park 904, 1098 XH Amsterdam, Netherlands}

 \emailAdd{oliver.buchmueller@cern.ch}
 \emailAdd{mdolan@slac.stanford.edu}
  \emailAdd{smalik@cern.ch}
 \emailAdd{c.mccabe@uva.nl}
\abstract{We introduce a Minimal Simplified Dark Matter (MSDM) framework to quantitatively characterise dark matter (DM) searches at the LHC. We study two MSDM models where the DM is a Dirac fermion which interacts with a vector and axial-vector mediator. The models are characterised by four parameters: $\mDM,\, \mMed,\, \gDM$ and~$\gq$, the DM and mediator masses, and the mediator couplings to DM and quarks respectively. The MSDM models accurately capture the full event kinematics, and the dependence on all masses and couplings can be systematically studied. The interpretation of mono-jet searches in this framework can be used to establish an equal-footing comparison with direct detection experiments. For theories with a vector mediator, LHC mono-jet searches possess better sensitivity than direct detection searches for light DM masses $(\lesssim5~\mathrm{GeV}$). For axial-vector mediators, LHC and direct detection searches generally probe orthogonal directions in the parameter space. We explore the projected limits of these searches from the ultimate reach of the LHC and multi-ton xenon direct detection experiments, and find that the complementarity of the searches remains. Finally, we provide a comparison of limits in the MSDM and effective field theory (EFT) frameworks to highlight the deficiencies of the EFT framework, particularly when exploring the complementarity of mono-jet and direct detection searches.
}

\begin{document}
\maketitle
\flushbottom

\section{Introduction}
\label{sec:intro}
Since the start-up of the Large Hadron Collider (LHC) in 2010, searches for direct particle dark matter (DM) production at colliders and their comparison with direct detection experiments have become a focal point for both the experimental and theoretical particle and astroparticle communities. 

In the past, searches for DM production at colliders~\cite{ATLAS:2012ky,ATLAS:2012zim,Chatrchyan:2012me,CMS-PAS-EXO-12-048}, including signatures with missing transverse energy (MET) such as mono-jets and monophotons, were quantitatively compared with results from direct detection experiments in the framework of an effective field theory (EFT). In this case, bounds are placed on the contact interaction scale $\Lambda$ of various higher dimensional operators describing the interaction of DM with the Standard Model fields~\cite{Cao:2009uw,Beltran:2010ww,Goodman:2010yf,Bai:2010hh,Goodman:2010ku,Rajaraman:2011wf,Fox:2011pm}.  While this  is in principle a model-independent way of interpreting these searches, the EFT approach fails severely in a number of circumstances making it an inappropriate framework to interpret DM searches at colliders. As shown in a previous paper~\cite{Buchmueller:2013dya} by some of the present authors, the EFT ansatz is only valid for a heavy mediator for which the mediator width is larger than its mass, making a particle-like interpretation of the mediator doubtful. Furthermore, for lighter mediator masses the EFT approach provides constraints which are either over-conservative (because the process is resonantly enhanced) or too aggressive (because the missing energy distribution is too soft). Further work discussing inadequacies of the EFT approach can be found in~\cite{Bai:2010hh,Fox:2011fx,Fox:2011pm,Fox:2012ru,Goodman:2011jq,Shoemaker:2011vi,Busoni:2013lha,Busoni:2014sya,Busoni:2014haa}, and these arguments also extend to searches for other mono-objects such as mono-photons~\cite{Fox:2011fx}, mono-leptons~\cite{Bai:2012xg} and mono-Higgs~\cite{Carpenter:2013xra,Petrov:2013nia}.

As  in~\cite{Buchmueller:2013dya}, a more appropriate approach to characterise DM searches at colliders is the use of simplified models~\cite{Alves:2011wf} (a similar approach was also advocated in~\cite{deSimone:2014pda}). This framework has proven to be very successful in searches for supersymmetry (SUSY) at the LHC~\cite{Aad:2013wta,Chatrchyan:2014lfa} and is by now the standard by which MET searches are compared in that context. Today, almost every ATLAS and CMS MET analysis provides an interpretation of their results in one or even several simplified models, which characterise the topologies probed by the search. 
 Although simplified models usually do not represent a full theory, valid constraints on more complete models such as the MSSM can be inferred if the simplified model approach is used with appropriate care~\cite{Buchmueller:2013exa,Barnard:2014joa,Papucci:2014rja,Kraml:2013mwa}.

Similarly, the advantage of the simplified model approach in the context of DM searches is that the full event kinematics over the whole parameter space is accurately captured. This allows for comprehensive studies of individual DM production topologies and for the optimisation of the experimental searches over all of the parameter space. In addition, the interpretation of collider searches in this framework can also be used to establish an equal-footing comparison with the results of direct detection experiments. This was not possible within the EFT approach where the collider limits were not valid in a large region of the relevant parameter space. 

In this paper we suggest a minimal set of simplified models that can be used to characterise DM searches both at colliders as well as direct detection experiments.  The parameter space of these models is defined by four variables: $\mMed$, $\mDM$, $\gq$, and $\gDM$, where $\mMed$ and $\mDM$ are the masses of the mediator and the DM particle, while $\gq$ and $\gDM$ represent the coupling of the SM particles to the mediator and the coupling of the mediator to the~DM. We use this minimal set of parameters to elucidate the true complementarity of mono-jet and direct detection searches. Continuing from~\cite{Buchmueller:2013dya}, we consider vector and axial- vector mediators to study the reach of the LHC and direct detection search from the LUX experiment. The mediator is produced in the s-channel at the LHC and in the t-channel at LUX (see figure~\ref{fig:OP}). 

We focus on the development of a consistent and state-of-the-art framework within which collider limits and direct detection limits can be interpreted and compared in a general way. Other collider searches, such as jets plus MET~\cite{Fox:2012ee} or di-jet searches~\cite{An:2012va}, as well as DM indirect detection searches~\cite{Ackermann:2013uma} will also possess significant sensitivity in constraining the DM parameter space defined by our simplified model approach. The inclusion of these other searches in our framework is relatively straightforward and will be the subject of future work. In addition, it is also possible to extend our approach to include t-channel mediators, and DM particles and mediators with different spin. Describing DM interactions with a scalar or pseudoscalar mediator requires additional technical details~\cite{Haisch:2012kf} that we leave for follow-up work.

This paper is structured as follows: in section~\ref{sec:SMS} we define our Minimal Simplified Dark Matter (MSDM) framework in the context of s-channel vector and axial-vector mediators. Section~\ref{sec:valid} contains technical details related to the CMS mono-jet search (including a discussion on the optimal MET cut in the MSDM interpretation) and direct detection experiments. Section~\ref{sec:results} contains our main results: we show the current complementarity of mono-jet and direct detection searches for vector and axial-vector mediators in various two-dimensional projections of the four-dimensional parameter space. We also have a dedicated discussion (in section~\ref{sec:low}) of the low mass region where direct detection experiments lose sensitivity and show projected limits (in section~\ref{sec:pro}) from future scenarios, including limits from the LHC after $30~\mathrm{fb}^{-1}$, $300~\mathrm{fb}^{-1}$ and~$3000~\mathrm{fb}^{-1}$ and xenon direct detection experiments with multiple ton-year exposures. In section~\ref{sec:compEFTMSDM} we present a comparison of the limits obtained in the MSDM and EFT frameworks, which serves to highlight the inadequacies of the EFT framework. We present our conclusions in section~\ref{sec:conc}. 

\section{Minimal Simplified Dark Matter models}
\label{sec:SMS}

The use of simplified models to characterise new physics searches at the LHC has become a standard procedure in both the experimental and theoretical communities. The advantage of simplified models is that they are fully described by a small number of fundamental parameters, such as masses, couplings and/or cross-sections. All these parameters are directly related to experimental observables, making this approach an effective framework for characterising searches in a well-defined, simple, and consistent manner.

In this paper we introduce a Minimal Simplified Dark Matter (MSDM) framework, which extends the SM matter content by two new fields whose properties are specified by (a minimum of) four parameters. The two fields are the dark matter and the mediator while the four basic parameters are the mass of the dark matter particle, $\mDM$, the mass of the mediator, $\mMed$, the coupling of quarks to the mediator, $\gq$, and the coupling of the mediator to the dark matter, $\gDM$. This set of parameters is sufficient to characterise the interactions of a variety of different UV completions (which we assume do not interact with each other) of the effective operators previously considered in the context of mono-jet searches (see e.g.~\cite{Goodman:2010ku} for a comprehensive list), including both s-channel and t-channel mediators~\cite{Chang:2013oia,An:2013xka,Bai:2013iqa,DiFranzo:2013vra,Papucci:2014iwa,Hamaguchi:2014pja,Garny:2014waa}.

\begin{figure}[t!]
\centering
\includegraphics[width=0.99\columnwidth]{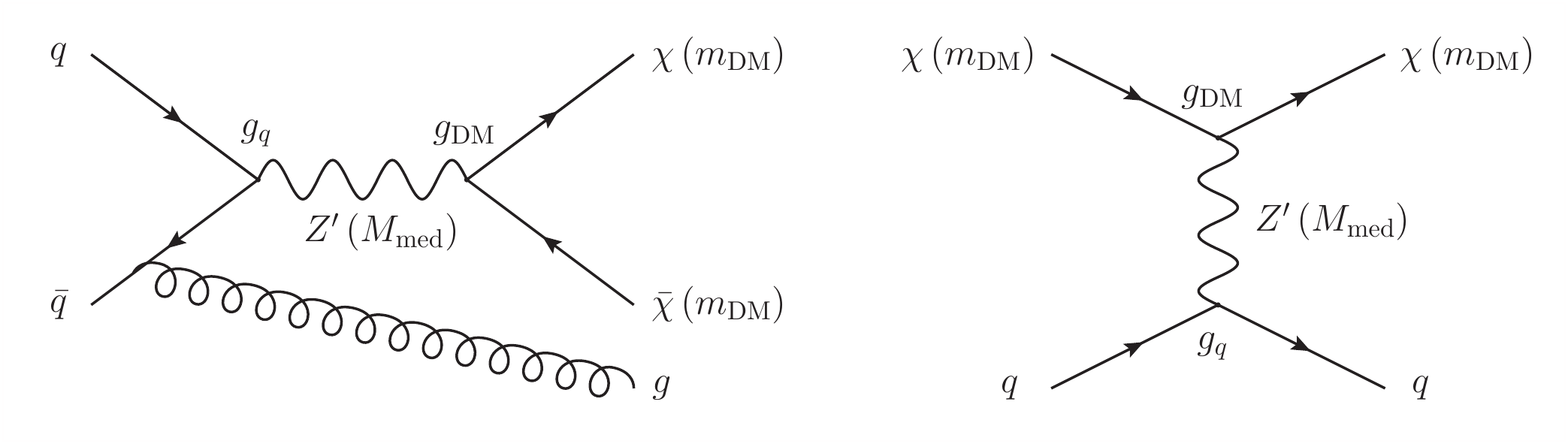}
\caption{The left diagram shows a contributing diagram for mono-jet production with an (axial) vector mediator at a hadron collider. The process is characterised by $\mMed ,\, \mDM ,\, \gDM$~and~$\gq$, which are the mediator and dark matter masses, and the mediator couplings to dark matter and quarks respectively. The right diagram shows the corresponding scattering process relevant for direct detection, which is characterised by the same four parameters.}
\label{fig:OP}
\end{figure}

In this paper we focus on the example of a vector mediator $Z'$ which is exchanged in the s-channel in mono-jet production. We consider the case when the dark matter is a Dirac fermion~$\chi$ and assume that the quark-mediator coupling~$\gq$ is equal for all quarks. In this case, as shown schematically in figure~\ref{fig:OP}, the model is completely characterised by the four parameters discussed above. These parameters are sufficient to determine the mono-jet production and direct detection scattering rate.

In general a vector mediator can have vector or axial-vector couplings with quarks and the dark matter. In addition to the usual mass and kinetic terms for $\chi$ and $Z'$, our MSDM model with a vector mediator is defined by the interaction terms
\begin{align}
\label{eq:AV} 
\mathcal{L}_{\mathrm{vector}}&\supset-\sum_q \gq Z'_{\mu} \bar{q}\gamma^{\mu}q- \gDM Z'_{\mu} \bar{\chi}\gamma^{\mu}\chi \\
\mathcal{L}_{\rm{axial}}&\supset-\sum_q \gq Z'_{\mu} \bar{q}\gamma^{\mu}\gamma^5q- \gDM Z'_{\mu} \bar{\chi}\gamma^{\mu}\gamma^5\chi
\end{align}
for vector and axial-vector couplings respectively, where the sum extends over all quarks. Models of a vector~$Z'$ mediator in the context of collider and direct detection searches have also been discussed elsewhere in the literature~\cite{Fox:2008kb,Cassel:2009pu,Bai:2010hh,Fox:2011pm,Fox:2011qd,Mambrini:2011dw,Gondolo:2011eq,Frandsen:2011cg,Shoemaker:2011vi,Frandsen:2012rk,An:2012va,An:2012ue,Alves:2013tqa,Arcadi:2013qia,Lebedev:2014bba,Davidson:2014eia,Fairbairn:2014aqa,Soper:2014ska}. Although the collider phenomenology of the vector and axial-vector mediators is similar, at direct detection experiments they are very different. In the non-relativistic limit the vector interaction gives a spin-independent interaction that is coherently enhanced by the number of nucleons, while the axial-vector interaction gives a spin-dependent signal which is not. In general it is also possible to have mixed vector and axial-vector couplings (so that e.g.~the quarks have vector couplings while the dark matter has axial-vector couplings). At tree-level, the mixed interaction is spin-dependent and velocity-squared suppressed~\cite{Kumar:2013iva} (where $v_{\rm{DM}}\simeq10^{-3}c$) so the loop-level spin-independent contribution, which is not velocity suppressed, dominates~\cite{Crivellin:2014qxa}. A treatment of the loop-induced contribution is beyond the scope of this work so we do not consider the case of mixed vector and axial-vector couplings further.

As both hadron collider and direct detection searches for dark matter primarily probe the interactions of dark matter with quarks, we set the mediator interactions with leptons to zero; the lepton couplings play no role (at tree-level) in the phenomenology in either hadron collider and direct detection searches~\cite{Kopp:2009et,Bell:2014tta}. While setting the mediator couplings to leptons to zero often introduces anomalies into the theory~\cite{Beringer:1900zz}, this does not have to be the case~\cite{Fox:2011qd,Buckley:2011vc,Buckley:2011mm,Duerr:2013dza,Duerr:2013lka}. If leptonic mediator couplings are introduced, di-lepton resonance searches will provide further constraints on the space of MSDM models.

As has been discussed in the literature~\cite{Fox:2011pm,An:2012va,Buchmueller:2013dya}, the mediator width~$\Gamma_{\rm{med}}$ plays an important role in mono-jet searches. In our MSDM models, we calculate the width from the four free parameters in the simplified model. We assume that no additional visible or invisible decays contribute to $\Gamma_{\rm{med}}$ so that the total width is
\begin{equation}
\Gamma_{\rm{med}}\equiv\Gamma(Z'\to\bar{\chi}\chi)\Theta\left(\mMed-2\mDM\right)+\sum_q\Gamma(Z'\to\bar{q}q) \Theta\left(\mMed-2m_q\right)
\end{equation}
where the individual contributions for the vector and axial-vector cases are
\begin{equation}
\Gamma(Z'\to\bar{\chi}\chi)_{\rm{vector}}=\frac{\gDM^2 \mMed}{12\pi}\left(1+\frac{2 \mDM^2}{\mMed^2} \right)\sqrt{1-\frac{4 \mDM^2}{\mMed^2}}\label{eq:Gamma1}
\end{equation}
\begin{align}
\Gamma(Z'\to\bar{q}q)_{\rm{vector}}&= \frac{3 \gq^2 \mMed}{12\pi}\left(1+\frac{2 m_q^2}{\mMed^2} \right)\sqrt{1-\frac{4 m_q^2}{\mMed^2}}\\
\Gamma(Z'\to\bar{\chi}\chi)_{\rm{axial}}&=\frac{\gDM^2 \mMed}{12\pi} \left(1-\frac{4 \mDM^2}{\mMed^2}\right)^{3/2}\\
\Gamma(Z'\to\bar{q}q)_{\rm{axial}}&= \frac{3 \gq^2 \mMed}{12\pi}\left(1-\frac{4 m_q^2}{\mMed^2}\right)^{3/2}\label{eq:Gamma4}\;.
\end{align}
It is straightforward to incorporate additional visible or invisible contributions to $\Gamma_{\rm{med}}$ but we do not consider that here.

Finally, while in this article we fix the mediator couplings to all quarks $\gq$ to be equal, we do not enforce that this must be equal to the mediator couplings to dark matter $\gDM$. 
Having different values for $\gDM$ and $\gq$ is neither unreasonable nor unexpected: for instance, in the Standard Model there is a factor of six difference between the hypercharge couplings of the $\mathrm{SU}(2)_{\rm{L}}$ quark doublet and the right-handed electrons.

We end this section by again emphasising that our motivation here is to flesh out the framework of MSDM models so that the search results from collider and direct detection experiments can be accurately characterised. These minimal simplified models enable us to infer qualitative and quantitative properties of more complete DM models and allow a comparison of collider and direct detection experiments on equal footing. Variations of the assumptions which we have made in this paper should be explored in the future to fully map out the MSDM landscape; the approach should be extended to include, for instance, other spins for the mediator and dark matter particle, different coupling structures and additional experimental constraints.

\section{Experimental details and validation}
\label{sec:valid}

This section describes our technical implementation of the CMS mono-jet and LUX direct detection searches. Our limits agree with the respective collaboration's limits 
to 10\% or better for both the mono-jet search and the LUX results, which is fully sufficient for our use-case. In addition, we check that the~MET ($\MET$) cut used by the mono-jet analysis, which was optimised assuming the EFT framework, is also appropriate within the MSDM framework. We also provide the prescription for translating between scattering cross-sections, on which the direct detection community quote limits, and $\mMed ,\, \mDM ,\, \gDM$ and~$\gq$, the parameters of interest for our MSDM models.

\subsection{LHC simulation details}

We generate LHC events for both the EFT and the MSDM using an extension of the POWHEG BOX described in~\cite{Haisch:2013ata,Nason:2004rx,Frixione:2007vw,Alioli:2010xd}. This allows us to generate the process where DM particles are pair produced in association with a parton from the initial state, resulting in a mono-jet signature. The implementation of this process in POWHEG allows for the generation of the signal to next-to-leading-order (NLO) accuracy and for this to be matched consistently with a parton shower. As shown in~\cite{Haisch:2013ata}, including NLO corrections results in a small enhancement of the cross-section compared to leading-order (LO). However, more importantly, it also leads to a substantial reduction in the dependence on the choice of the renormalisation and factorisation scales and hence the theoretical uncertainty on the signal prediction. Generating the signal to NLO accuracy should therefore lead to more robust bounds. 
For the purposes of validating our framework against the results from the CMS mono-jet search~\cite{CMS-PAS-EXO-12-048}, events for the EFT process are generated at LO to enable a direct comparison with the limits provided by CMS. This is another advantage of  implementing this process in the POWHEG BOX since it is capable of simulating both the EFT case (where the mediator is integrated out) and the MSDM case (where the mediator is correctly taken into account). For this validation process, we follow CMS and use the CTEQ6L1~\cite{Pumplin:2002vw} parton distribution functions (PDFs).  
 
We set the renormalisation and factorisation scales to
\begin{equation}
\mu =  \sqrt{m_{\chi\bar{\chi}}^{2} + p_{T,j_{1}}^{2}} + p_{T,j_{1}}
\end{equation}
where $m_{\chi\bar{\chi}}$ is the invariant mass of the DM pair and $p_{T,j1}$ denotes the transverse momentum of the leading jet $j_1$. As noted in~\cite{Haisch:2013ata}, this choice of scale leads to NLO corrections that are relatively independent of $\mDM$. For all cases except the validation of the CMS EFT limits, we use the MSTW2008NLO~\cite{Martin:2009iq} PDFs. 

The parton level process produced by POWHEG is matched to \texttt{Pythia}~8.180~\cite{Sjostrand:2007gs} for showering and hadronisation and put through a detector simulation using \texttt{Delphes3}~\cite{Ovyn:2009tx,deFavereau:2013fsa}, with parameters that are tuned to the CMS detector. Subsequently, the mono-jet selection cuts described in~\cite{CMS-PAS-EXO-12-048} are applied: Jets are reconstructed using the anti-$k_t$~\cite{Cacciari:2008gp} algorithm with a distance parameter of 0.5. Events are selected where the highest $p_T$ jet has transverse momentum above 110 GeV and $|\eta_{j_1}| < 2.4$. Another jet is allowed if it has $p_T > 30$ GeV and $|\eta| < 4.5$, but an event is vetoed if there are any additional jets satisfying this requirement. Thus, the mono-jet signature comprises of either one or two-jet events. If there are two jets in the event, it is further vetoed if the angular separation in azimuth between the two jets,  $\Delta \phi_{j_1,j_2}$, is greater than 2.5. Seven signal regions are defined, with increasing thresholds for the missing transverse energy: $\MET > 250,\,300,\,350,\,400,\,450,\,500$ and 550~GeV. The CMS mono-jet analysis optimised the $\MET$ threshold by benchmarking it against exclusion sensitivity in the EFT interpretation framework. They found that the optimal threshold was 400 GeV. 
As discussed in section~\ref{sec:optimise}, we have verified that this threshold is also a reasonable choice for placing limits in the MSDM models so we use it throughout this paper in our mono-jet analysis.

\begin{figure}[t!]
\centering
\includegraphics[width=0.495\columnwidth]{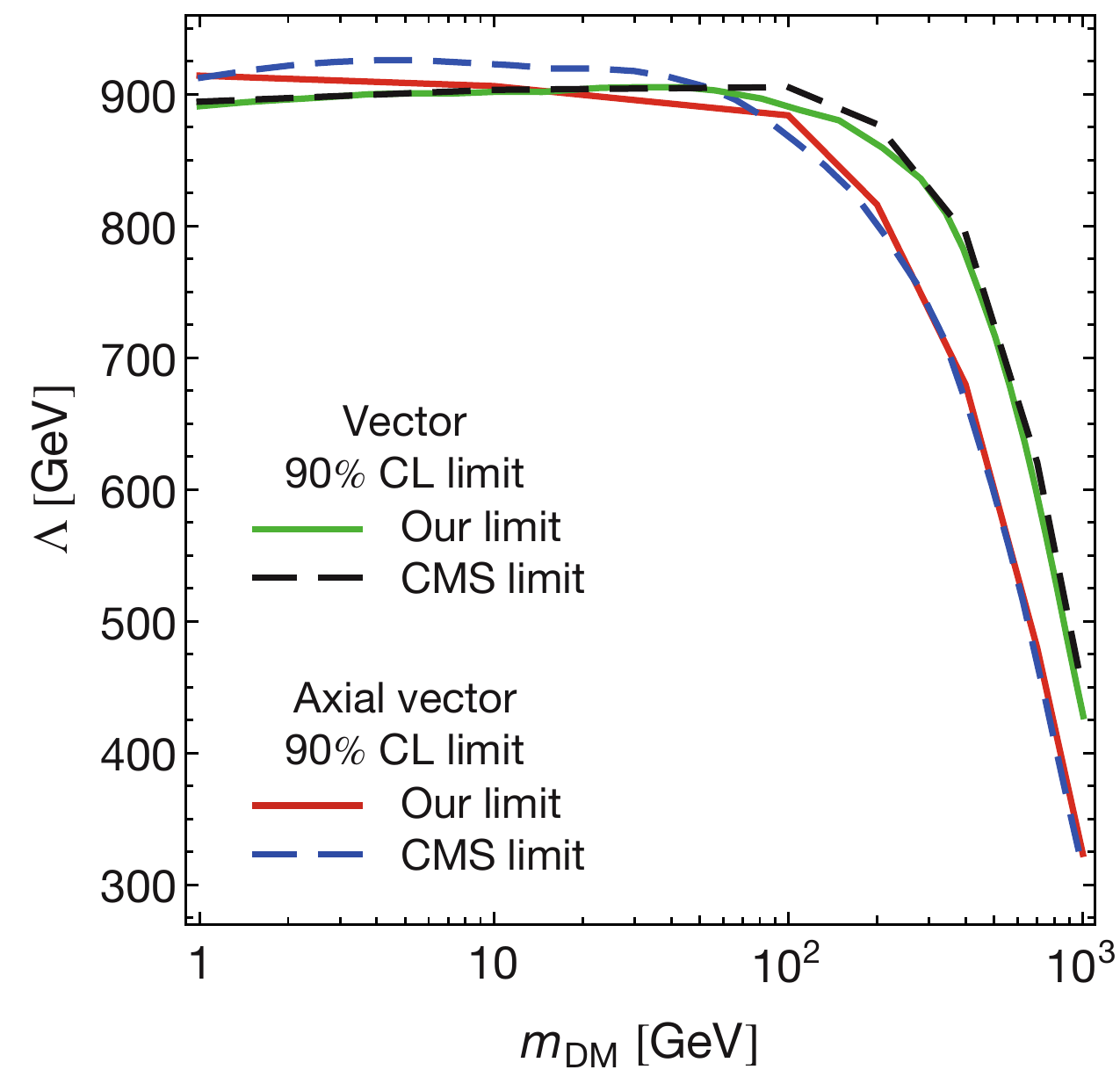} 
\includegraphics[width=0.495\columnwidth]{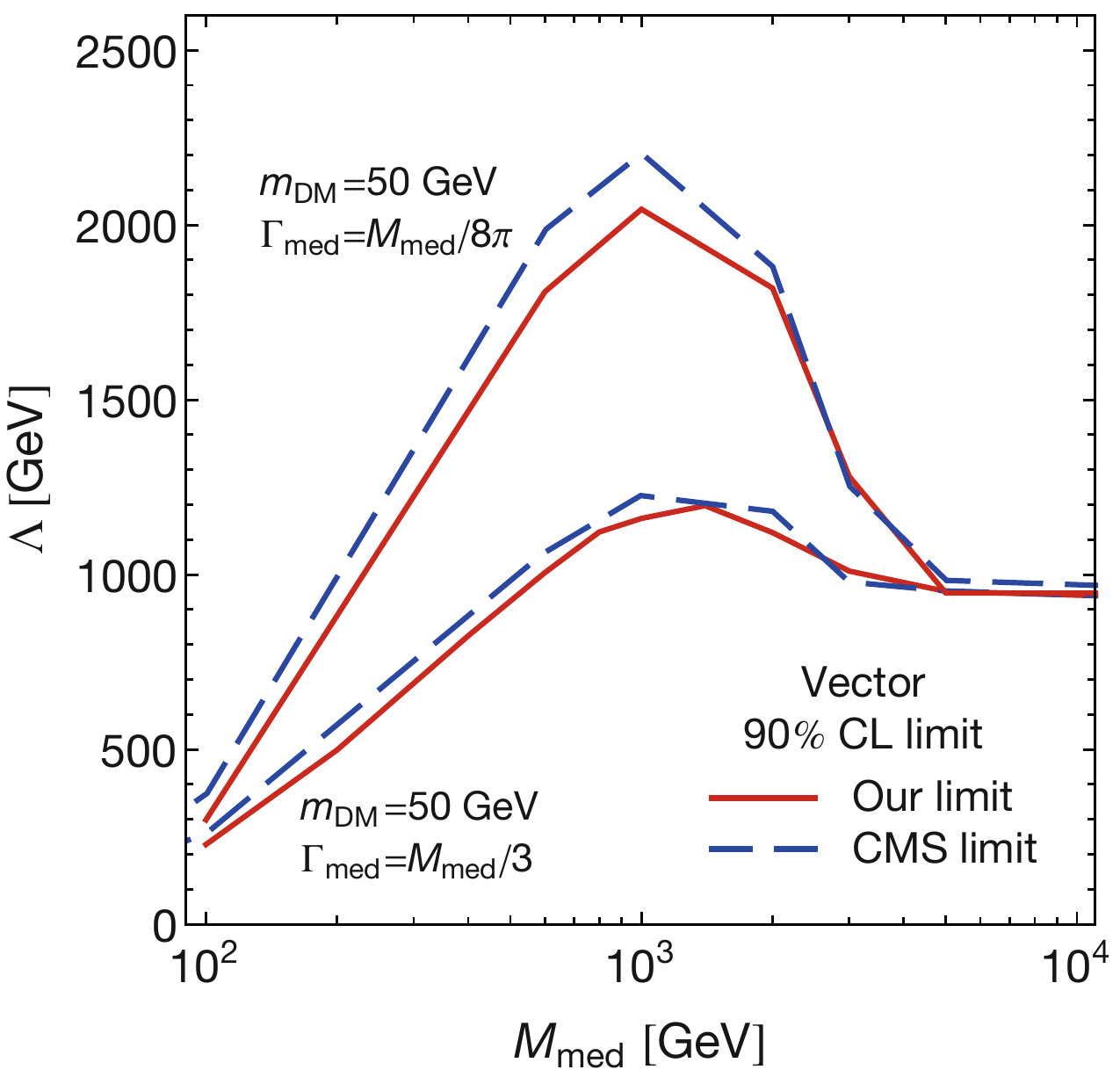}
\caption{A comparison of our 90\%~CL limits (solid lines) and the CMS 90\%~CL limits (dashed lines) on the contact interaction scale $\Lambda$. The left panel shows the results for the vector and axial-vector operators; our limits and the CMS limits agree to better than 5\%. The right panel shows the results for the vector interaction with two choices of the mediator width $\Gamma_{\rm{med}}$; our limits and the CMS limits agree to better than 10\%.}
\label{fig:validation}
\end{figure}

Figure~\ref{fig:validation} shows a comparison of our 90\%~CL limits (solid lines) and the CMS 90\% CL limits from~\cite{CMS-PAS-EXO-12-048} (dashed lines) on the contact interaction scale~$\Lambda$. The contact interaction scale~$\Lambda$ is the mass scale that gives the correct dimension to the vector and axial-vector higher dimensional operators
\begin{align}
\label{eq:EFT:vec}
\mathcal{L}^{\mathrm{EFT}}_{\mathrm{vector}}&\supset\sum_q \frac{1}{\Lambda^2} \bar{q}\gamma_{\mu}q\,\bar{\chi}\gamma^{\mu}\chi\\
\mathcal{L}^{\mathrm{EFT}}_{\rm{axial}}&\supset\sum_q\frac{1}{\Lambda^2} \bar{q}\gamma_{\mu}\gamma^5q\,\bar{\chi}\gamma^{\mu}\gamma^5\chi\;,
\label{eq:EFT:axvec}
\end{align}
where the sum extends over all quarks. The left panel shows the results for the vector (green and black lines) and axial-vector (red and blue lines) operators while the right panel shows the results for the vector operator for two choices of $\Gamma_{\rm{med}}$. Our limits and the CMS limits agree within 5\% (10\%) in the left (right) panel. We, therefore, conclude that we are able to reproduce the CMS mono-jet analysis to 10\% or better, which is fully sufficient for our use-case. 

\begin{figure}[t!]
\begin{center}
\hspace{8mm}\includegraphics[width=0.65\columnwidth]{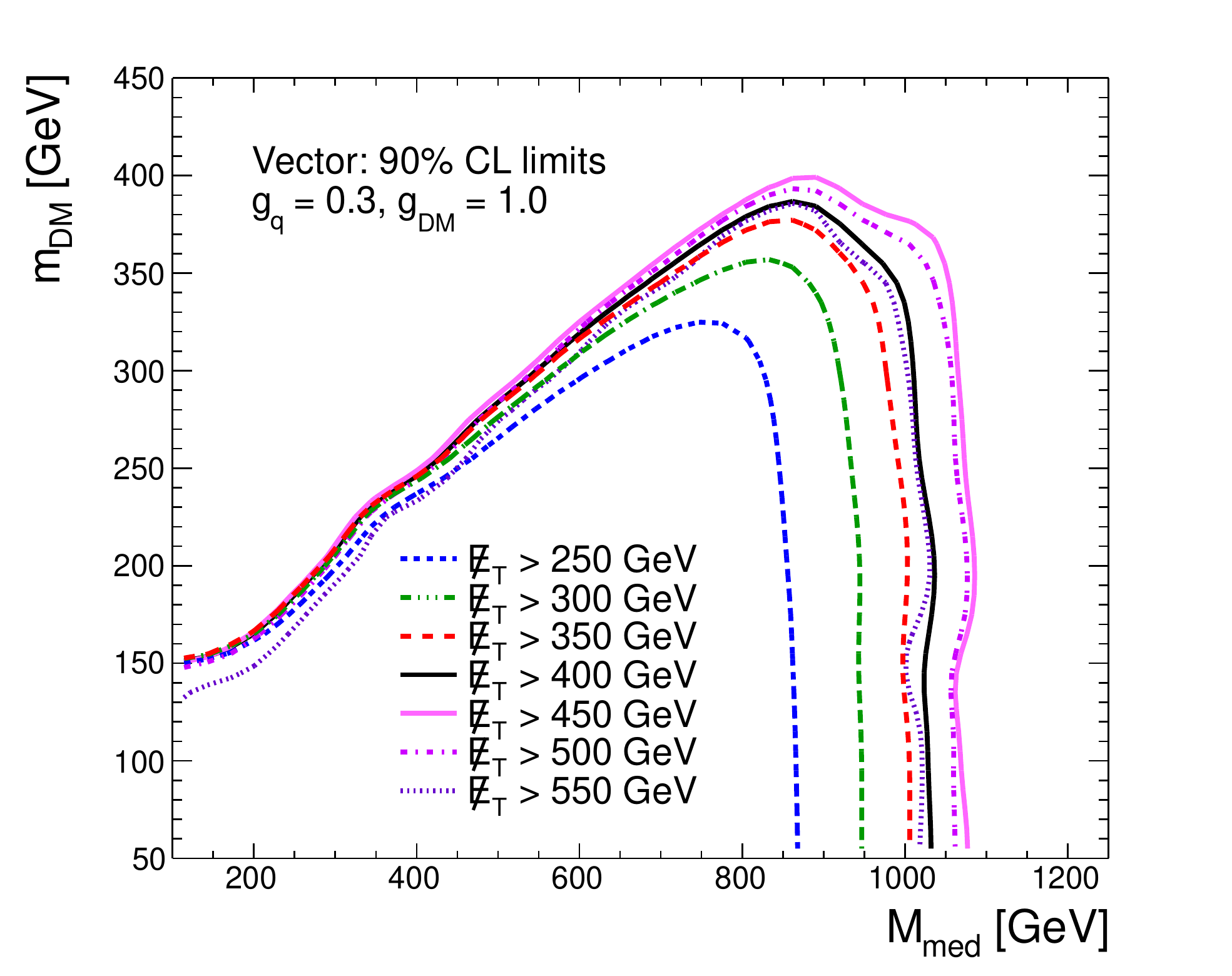}
\caption{Exclusion limits at 90\%~CL for the vector mediator in the $\mDM$-$\mMed$ plane for various different values of the $\MET$ threshold. The couplings are fixed to $g_q = 0.3,\,\gDM = 1$. The optimum $\MET$ cut is~$\MET>450$~GeV, slightly larger than the $\MET>400$~GeV cut used in the CMS (and our)  analysis. The 50~GeV increase in the $\MET>450$~GeV exclusion contour is modest so the $\MET>400$~GeV cut is a reasonable choice.}
\label{fig:multibin}
\end{center}
\end{figure}

\subsection{Optimising mono-jet searches in the MSDM framework}
\label{sec:optimise}

Fully exploiting the adoption of simplified models as a framework for the presentation and interpretation of mono-jet searches requires a re-optimisation of current experimental search cuts, which are based on signal samples generated in the EFT framework. While carrying out a complete re-optimisation of all cuts is beyond the scope of this paper, we can use the information provided by CMS to briefly examine the effects of varying the $\MET$ cut on the limits set in our MSDM framework. The $\MET$ cut is the most important cut of this analysis to separate background from a potential signal.

In figure~\ref{fig:multibin} we show the reach in the vector MSDM model using the expected limits provided by CMS for different values of $\MET$ from 250 to 550~GeV. In this example the couplings are fixed to $g_q = 0.3,\,\gDM = 1$. We find that the optimum MET cut for the MSDM model is~$\MET>450$~GeV. This is slightly larger than the $\MET>400$~GeV cut which is optimal in the EFT framework. However, the 50~GeV increase in the exclusion contour is rather modest and thus we conclude that the use of the default cut of $\MET>400$~GeV is still reasonable.

\subsection{Facets of direct detection}
\label{sec:DD}

Dark matter interactions with nuclei lead to either spin-independent (SI) or spin-dependent (SD) scattering in the non-relativistic limit (recall that  in the galaxy $v_{\rm{DM}}\simeq10^{-3}c$). Therefore, limits on the cross-section to scatter off a nucleon are presented separately for each case. In this paper we primarily use the limits from the LUX experiment, which currently sets the strongest published limit on SI interactions for $\mDM\gtrsim6$~GeV. Two-phase xenon detector technology has a proven track record through the ZEPLIN~\cite{Alner:2005pa,Alner:2007ja,Akimov:2011tj}, XENON~\cite{Angle:2007uj,Aprile:2012nq,Aprile:2013doa} and LUX~\cite{Akerib:2013tjd} programs and is scalable to the much larger target masses required to probe very small scattering cross-sections. Xenon experiments also have the advantage that they are sensitive to both SI and SD interactions.

LUX has published the limit for SI interactions but not for SD interactions. We now describe our procedure for calculating both limits and assume that in the future, both limits will be provided directly by the collaboration.  We model the differential scattering rate per kg-day at LUX with~\cite{Davis:2012vy,Davis:2012hn}
\begin{equation}
\frac{dR}{d\mathrm{S1}}=\xi(m_{\mathrm{DM}})\int_{3\,\rm{keV}}^{\infty}dE_{\rm{R}}\sum_{n=1}^{\infty}\epsilon_{\mathrm{S1}}(\mathrm{S1})\epsilon_{\rm{S2}}(E_{\rm{R}}) \mathrm{N}(\mathrm{S1};n,\sqrt{n}\sigma_{\mathrm{PMT}})\ \mathrm{P}(n;\nu(E_{R}))\frac{dR}{dE_{\rm{R}}}\;,
\end{equation}
where $\epsilon_{\mathrm{S1}}(\mathrm{S1})$ and $\epsilon_{\rm{S2}}(E_{\rm{R}})$ are the S1 and S2 efficiencies~\cite{Akerib:2013tjd}, S1 ranges from 2~PE to 30~PE, $\mathrm{N}(\mathrm{S1};n,\sqrt{n}\sigma_{\rm{PMT}})$ is a Normal distribution with mean and variance $n$ and $n \sigma_{\rm{PMT}}^2$ respectively, where $\sigma_{\rm{PMT}}\approx0.37$~\cite{Akerib:2012ys}, $\mathrm{P}(n;\nu(E_{R}))$ is a Poisson distribution with mean $\nu(E_R)\equiv8.8\, E_{\rm{R}}\,\mathcal{L}_{\rm{eff}}(E_{\rm{R}})$, where the numerical pre-factor is quoted in~\cite{Akerib:2013tjd} and we use $\mathcal{L}_{\rm{eff}}(E_{\rm{R}})$ from NEST (at 181~kV/cm)~\cite{NEST2013a}.

Following the event selection used by the LUX collaboration, in our analysis we only consider events that fall below the mean of the nuclear recoil band; the efficiency factor $\xi(m_{\rm{DM}})$ gives the DM mass-dependent fraction of events that fall below this mean. We calculate it by simulating the distribution of DM events in the $\log_{10}(\mathrm{S2}_{\rm{b}}/\mathrm{S1})$--$\mathrm{S1}$ plane for different values of $\mDM$. We find that $\xi\approx1$ at $\mDM=5$~GeV, falling to $\xi\approx0.5$ at $\mDM=200$~GeV and remaining constant at 0.5 for higher masses.

The differential event rate for dark matter with local density $\rho_{\rm{DM}}=0.3~\text{GeV}\,\text{cm}^{-3}$ to scatter on a nucleus with mass $m_{\rm{N}}$ is~\cite{Lewin:1995rx}
\begin{equation}
\frac{dR}{dE_{\rm{R}}}=\frac{\rho_{\rm{DM}}}{m_{\rm{N}}\mDM}\int_{v_{\rm{min}}}^{\infty}v f(\mathbf{v},\mathbf{v}_{\rm{E}})\frac{d \sigma}{d E_{\rm{R}}}\,d^3v\;,
\label{eq:dRdE1}
\end{equation}
where $f(v)$ is the local dark matter velocity distribution in the galactic frame, $v=|\mathbf{v}|$, $\mathbf{v}_{\rm{E}}$ is the Earth's velocity relative to the galactic rest frame~\cite{Lee:2013xxa,McCabe:2013kea} and $v_{\rm{min}}$ is the minimum speed required for a nucleus to recoil with energy $E_{\rm{R}}$. Here, we use the usual benchmark values for the astrophysical parameters that are used in the direct detection community. Uncertainties in these parameters lead to an uncertainty of around 50\% on the 90\%~CL nucleon scattering cross-section limit~\cite{McCabe:2010zh,Fairbairn:2012zs}.

\begin{figure}[t!]
\centering
\includegraphics[width=0.495\columnwidth]{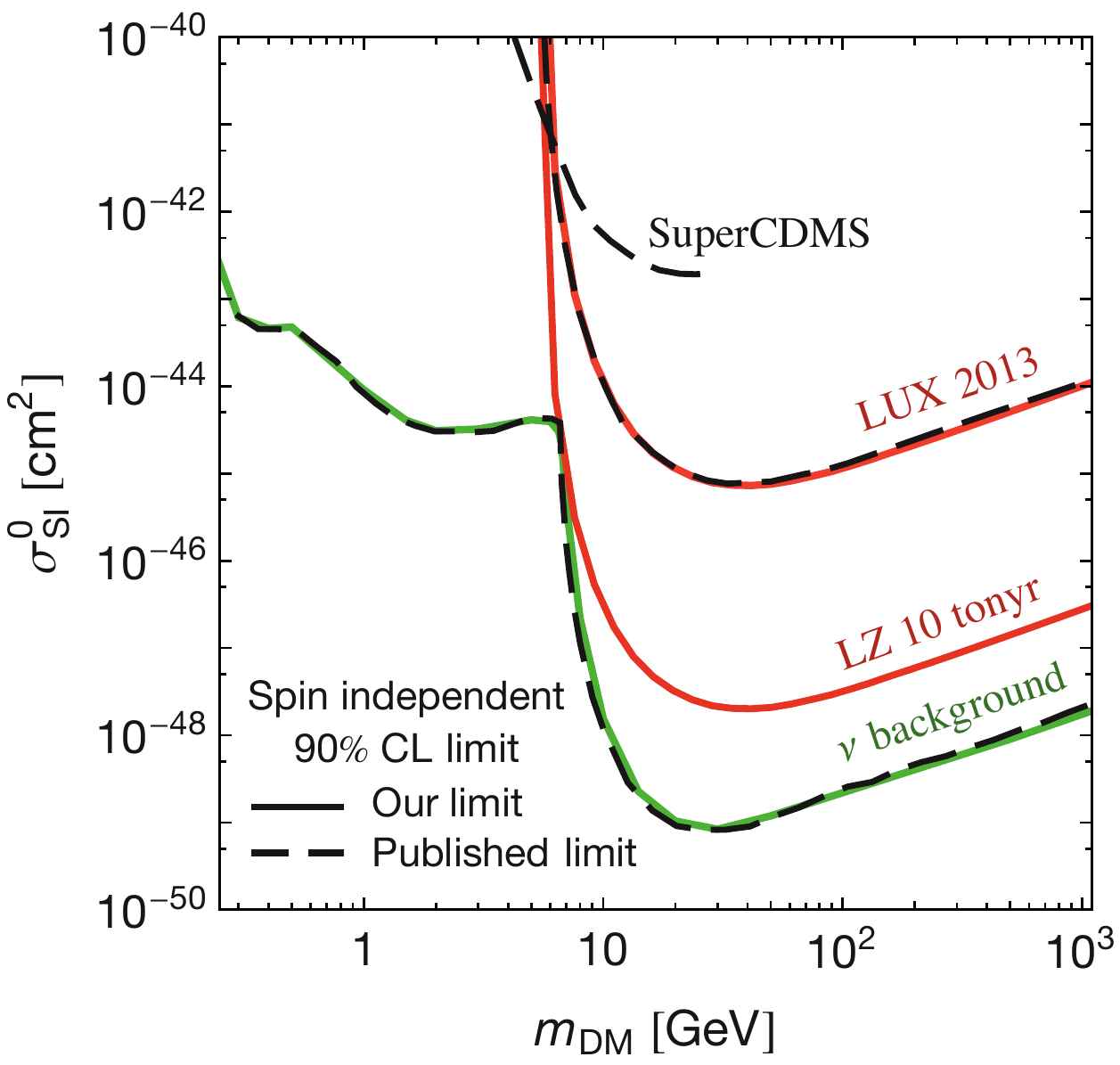} 
\includegraphics[width=0.495\columnwidth]{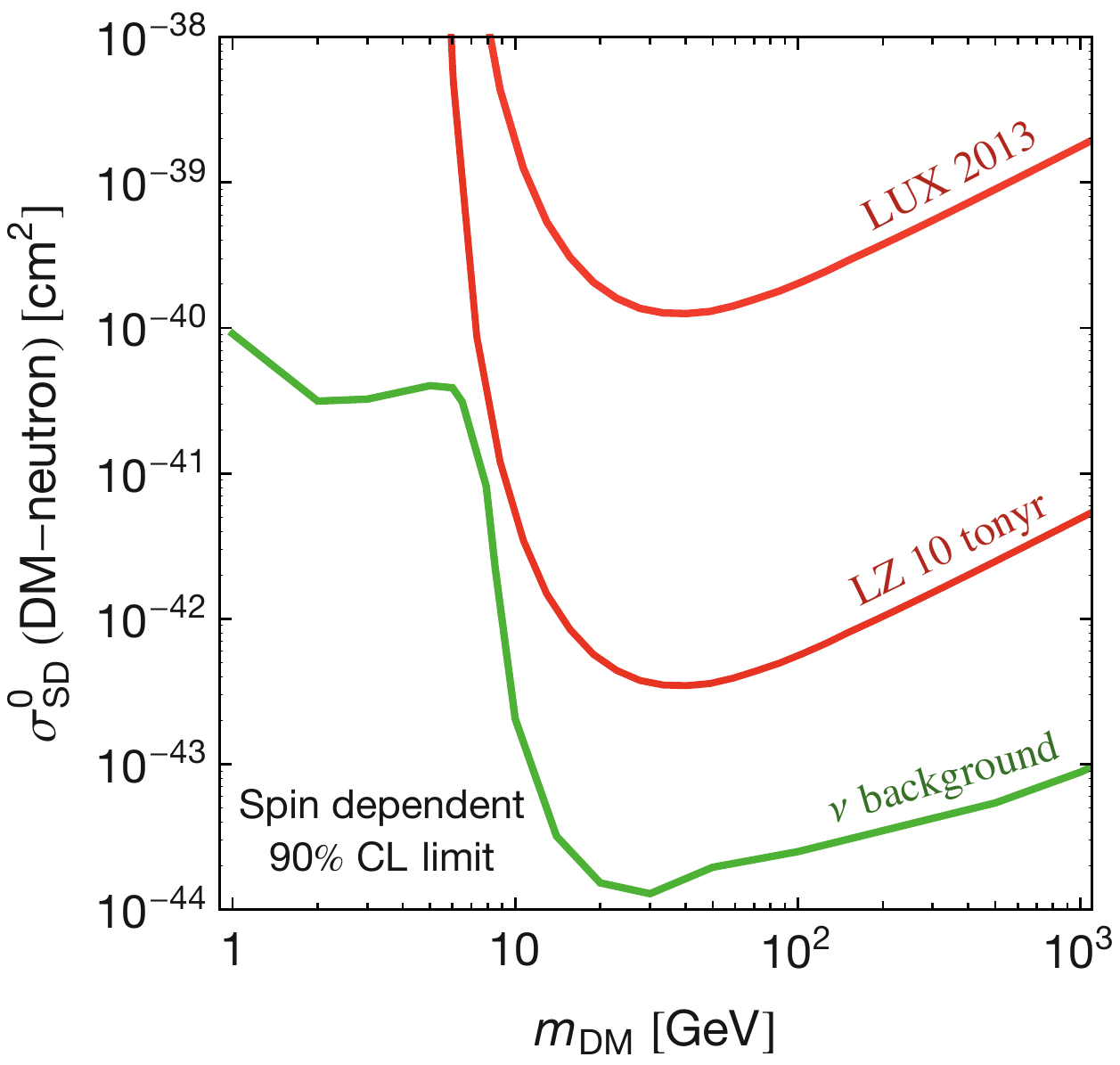}
\caption{The left panel shows a comparison of our 90\% CL limits (solid lines) and published 90\%~CL limits (dashed lines) on spin-independent direct detection cross-sections $\sigma^0_{\rm{SI}}$. The right panel shows our limits for the corresponding spin-dependent direct detection cross-sections $\sigma^0_{\rm{SD}}$, for which no published results are available to compare with. Also shown in both plots is the expected reach of LZ with a 10 ton-year exposure, and the discovery reach accounting for the irreducible neutrino background from~\cite{Billard:2013qya} (dashed line) and our calculation of the same limit (solid line).}   
\label{fig:DDvalidation}
\end{figure}

The input from particle physics (as opposed to astrophysics e.g.~the DM velocity distribution) enters through $d\sigma/dE_{\rm{R}}$, the differential cross-section to scatter off a nucleus. For SI and SD scattering, this is given by
\begin{align}
\frac{d \sigma_{\rm{SI}}}{d E_{\rm{R}}}&=\frac{m_{\rm{N}}\, A^2\, \sigma^0_{\rm{SI}}}{2 \mu^2_{n\chi} v^2}F^2(E_{\rm{R}})\label{eq:SI}\\
\frac{d \sigma_{\rm{SD}}}{d E_{\rm{R}}}&=\frac{4\pi}{3}\frac{m_{\rm{N}}\,\sigma^0_{\rm{SD}}}{2 \mu^2_{n\chi}v^2}\sum_i \frac{\mathfrak{f}_i}{2 J_i+1}S_i(E_{\rm{R}})\label{eq:SD}\;,
\end{align}
where $\mu_{n\chi}$ is the DM-nucleon reduced mass, $A\approx131$ is the atomic number of xenon, $F(E_{\rm{R}})$ is the Helm nuclear form factor~\cite{Duda:2006uk}, $S(E_{\rm{R}})$ is the spin structure function~\cite{Klos:2013rwa}, $J$ is the nucleus spin, the sum extends over all isotopes with non-zero spin and $\mathfrak{f}_i$ is the fractional abundance of the $i^{th}$ isotope with non-zero spin. The cross-sections $\sigma^0_{\rm{SI}}$ and $\sigma^0_{\rm{SD}}$ are the SI and SD cross-sections to scatter off a nucleon (in the limit of zero-momentum transfer) and are the quantities on which the experimental limits are quoted. The proton structure function is suppressed by over an order of magnitude relative to the neutron structure function for xenon isotopes therefore we ignore its contribution. This means that $\sigma^0_{\rm{SD}}$ is the cross-section to scatter off a neutron. Note that the dominant difference between the SI and SD cross-sections in eqs.~\eqref{eq:SI} and~\eqref{eq:SD} comes from the factor $A^2$. This enhancement ensures that the SI limits are over $10^4$ stronger than the corresponding SD limits, which plays an important role when comparing the LUX result with the mono-jet search (in section~\ref{sec:results}).

The left and right panels of figure~\ref{fig:DDvalidation} show the limits on $\sigma^0_{\rm{SI}}$ and $\sigma^0_{\rm{SD}}$ respectively. In order to validate our procedure we compare our $\sigma^0_{\rm{SI}}$ limit (solid lines) with the published ones (dashed lines).  Our LUX limits (red) are calculated with the `pmax' method, introduced in ~\cite{Yellin:2002xd} and the limit on $\sigma^0_{\rm{SI}}$ agrees with the published LUX limit to better than 10\% over the whole mass range. Based on this observation we assume that our limits for $\sigma^0_{\rm{SD}}$, for which no published limits from the LUX Collaboration are available, are also a good approximation of the performance of the experiment. The limit we find for $\sigma^0_{\rm{SD}}$ is a factor $\sim2.5$ stronger than the published XENON100 SD limit to scatter off a neutron~\cite{Aprile:2013doa}. This is the same relative improvement as found in the SI limits of LUX and XENON100. As expected, the limit on $\sigma^0_{\rm{SD}}$ is significantly weaker than the limit on $\sigma^0_{\rm{SI}}$. 

Also shown is the limit on~$\sigma^0_{\rm{SI}}$ from SuperCDMS~\cite{Agnese:2014aze}, which is stronger than LUX's for $\mDM\lesssim6$~GeV, the projected limits from LZ (the successor to LUX) assuming an exposure of 10 ton-years~\cite{Malling:2011va}, and the discovery reach when coherent neutrino scattering is taken into account~\cite{Billard:2013qya}. For the LZ limit we assume that the efficiencies and background event rate remain the same as at LUX. The discovery reach indicates the cross-section at which 90\% of experiments can make a $3\sigma$ discovery of dark matter, taking into account the background contribution from coherent neutrino scattering. The result in~\cite{Billard:2013qya} is for SI only, therefore we reproduce that result here in the left panel of figure~\ref{fig:DDvalidation} (finding good agreement over the whole mass range) and extend the calculation to calculate the limit for SD scattering in the right panel of figure~\ref{fig:DDvalidation}. We make the same assumptions in our calculation as those used in~\cite{Billard:2013qya}. Namely, we assume that the low mass reach is from a xenon experiment with an exposure of 200~kg-years and an energy threshold of 3~eV, while the high mass reach is from a xenon experiment with a 4~keV energy threshold and the exposure is such that 500 neutrino events are expected. Furthermore, the magnitude, distribution and uncertainty of the neutrino fluxes are the same as those in~\cite{Billard:2013qya} and we assume that the experimental efficiencies and energy resolution is perfect, and we ignore the effect of neutrinos scattering on electrons. We make use of these results in sections~\ref{sec:low} and~\ref{sec:pro}.
 Finally, we note that the discovery reach may be improved with new experimental and theoretical techniques (see e.g.~\cite{Grothaus:2014hja,Ruppin:2014bra}) but we do not discuss that further here. 

Translating the limits on $\sigma^0_{\rm{SI}}$ and $\sigma^0_{\rm{SD}}$ to constraints on the parameters of interest in our MSDM models is now straightforward. For the vector model, the scattering interaction is SI and the cross-section to scatter off a point-like nucleon in the non-relativistic limit is
\begin{align}
\label{eq:SI2MSDM}
\sigma^0_{\rm{SI}}&=\frac{9\, g^2_{\rm{DM}}\,g_q^2\, \mu^2_{n\chi}}{\pi M^4_{\rm{med}}}\\
&\approx1.1\times 10^{-39}~\mathrm{cm}^2\cdot\left(\frac{\gDM \,g_q}{1}\right)^2\left( \frac{1~\mathrm{TeV}}{M_{\rm{med}}}\right)^4\left(\frac{\mu_{n\chi}}{1~\mathrm{GeV}} \right)^2\;.
\end{align}
In the axial-vector model, the scattering interaction is SD and the analogous result for the cross-section to scatter off a point like neutron is
\begin{align}
\label{eq:SD2MSDM}
\sigma^0_{\rm{SD}}&=\frac{3\, g^2_{\rm{DM}}\,g_q^2 (\Delta_u+\Delta_d+\Delta_s)^2\, \mu^2_{n\chi}}{\pi M^4_{\rm{med}}}\\
&\approx 4.6\times 10^{-41}~\mathrm{cm}^2\cdot\left(\frac{\gDM \,g_q}{1}\right)^2\left( \frac{1~\mathrm{TeV}}{M_{\rm{med}}}\right)^4\left(\frac{\mu_{n\chi}}{1~\mathrm{GeV}} \right)^2\;,
\end{align}
where $\Delta_u=-0.42$, $\Delta_d=0.85$ and $\Delta_s=-0.08$~\cite{Cheng:2012qr}. Other values for $\Delta_u$, $\Delta_d$ and $\Delta_s$ are also used in the literature (see e.g.~\cite{Ellis:2000ds,Gondolo:2004sc,Belanger:2008sj}) and differ by~$\mathcal{O}(10\%)$. We assume that the coupling~$g_q$ is equal for all quarks in both of these results. Note that the dependence on $\mDM$ in eqs.~\eqref{eq:SI2MSDM} and~\eqref{eq:SD2MSDM} essentially vanishes for $\mDM\gtrsim10$~GeV, since $\mDM$ enters only through the reduced mass~$\mu_{n\chi}$. For large values of $\mDM$, the dark matter mass only enters the scattering rate through its explicit dependence in eq.~\eqref{eq:dRdE1}. As a result (and as can be verified from figure~\ref{fig:DDvalidation}) the limits on $\sigma^0_{\rm{SI}}$ and $\sigma^0_{\rm{SD}}$ scale in proportion to $\mDM$ at large $\mDM$. Finally, we note that eqs.~\eqref{eq:SI2MSDM} and~\eqref{eq:SD2MSDM} are valid when the mediator mass is greater than the typical momentum transfer in the scattering process, which is $\sim100$~MeV. Therefore in the following, we do not consider direct detection limits for $M_{\rm{med}}<100$~MeV.

\section{Landscape of collider and direct detection searches}
\label{sec:results}

In this section we present our 90\% exclusion limits from the LHC mono-jet and LUX searches for the MSDM models defined in section~\ref{sec:SMS}. We discuss the vector and axial-vector cases in parallel, showing first the current constraints on the whole parameter space (in section~\ref{sec:current}) before zooming in on the low $\mDM$ region where direct detection searches start to lose sensitivity (in section~\ref{sec:low}). We conclude this section with projections for future searches (in section~\ref{sec:pro}).

To explore the complementarity of collider and direct detection searches we map out the four-dimensional parameter space of the MSDM models by showing projections in two parameters. The following two-dimensional planes are considered:   

\begin{itemize}
\item $\mDM$ vs $\mMed$, for fixed values of couplings $g_q$ and $\gDM$, both for the case where $g_q = \gDM$ and $g_q \neq \gDM$.
\item  $\mMed$ vs ($g_{q}=\gDM$), for fixed values of $\mDM$.
\item  $\mDM$ vs ($g_{q}=\gDM$), for fixed values of $\mMed$.
\item  $\gDM$ vs $g_q$, for fixed values of $\mDM$ and $\mMed$.
\end{itemize}

We limit the parameter space by the requirement that the mediator width $\Gamma_{\rm{med}}$ is not larger than its mass $\mMed$. From eqs.~\eqref{eq:Gamma1} to~\eqref{eq:Gamma4}, we find that for the case $\gDM =g_q$, this confines the maximal value of the two couplings $\gDM =g_q \approx 1.45$, when the mediator couples to all quarks and $\mMed>2 m_t$.  For a few plots we slightly extend beyond this coupling value to better visualise the limits.  For all figures which show a direct comparison of the LHC and LUX searches, the axial-vector figures have linear scales while the vector figures have log scales. This is to better display the different features of collider and direct detection sensitivities.   

\subsection{Limits from current searches}
\label{sec:current}

\begin{figure}[t!]
\centering
\includegraphics[width=0.495\columnwidth]{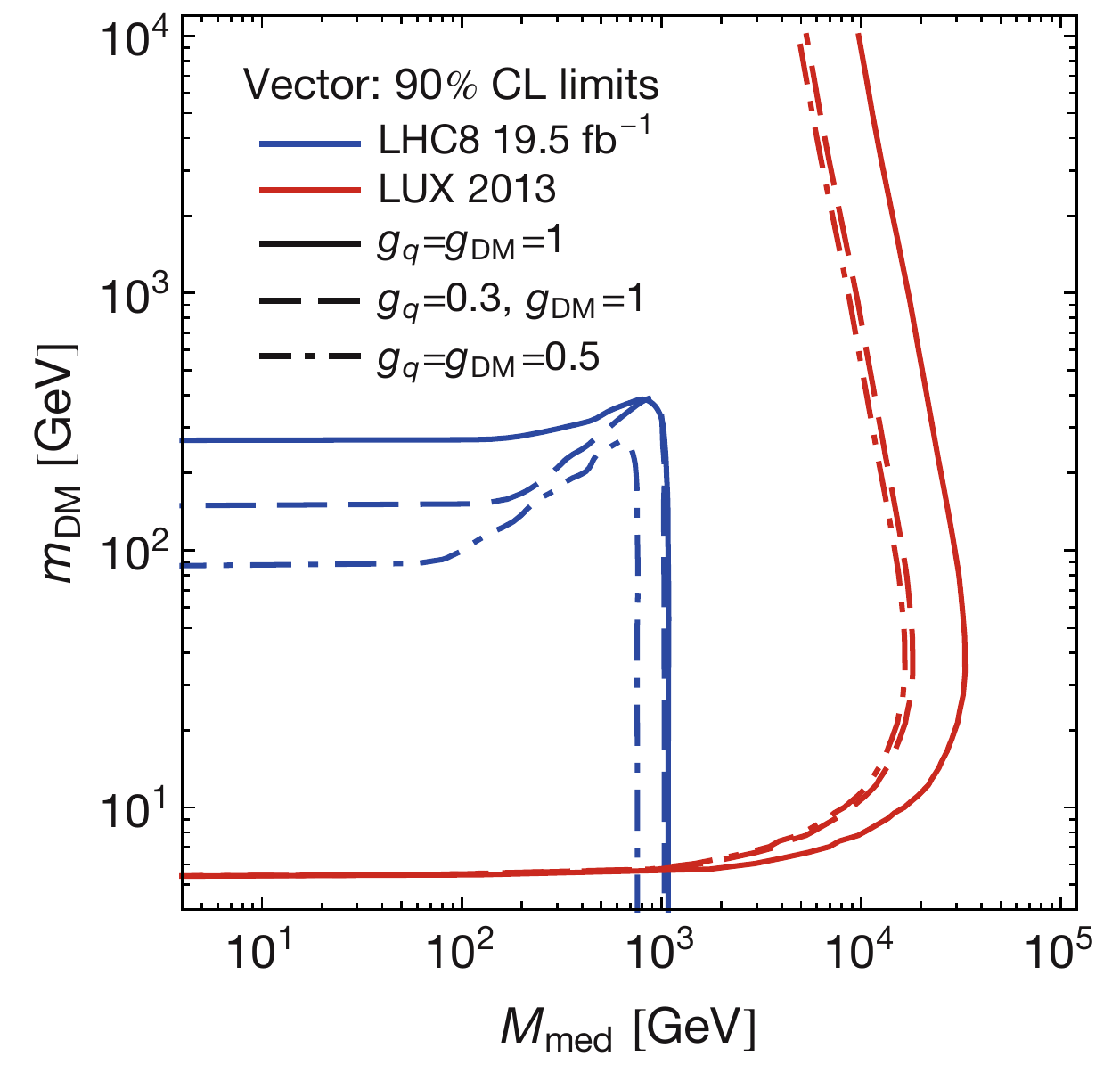}
\includegraphics[width=0.495\columnwidth]{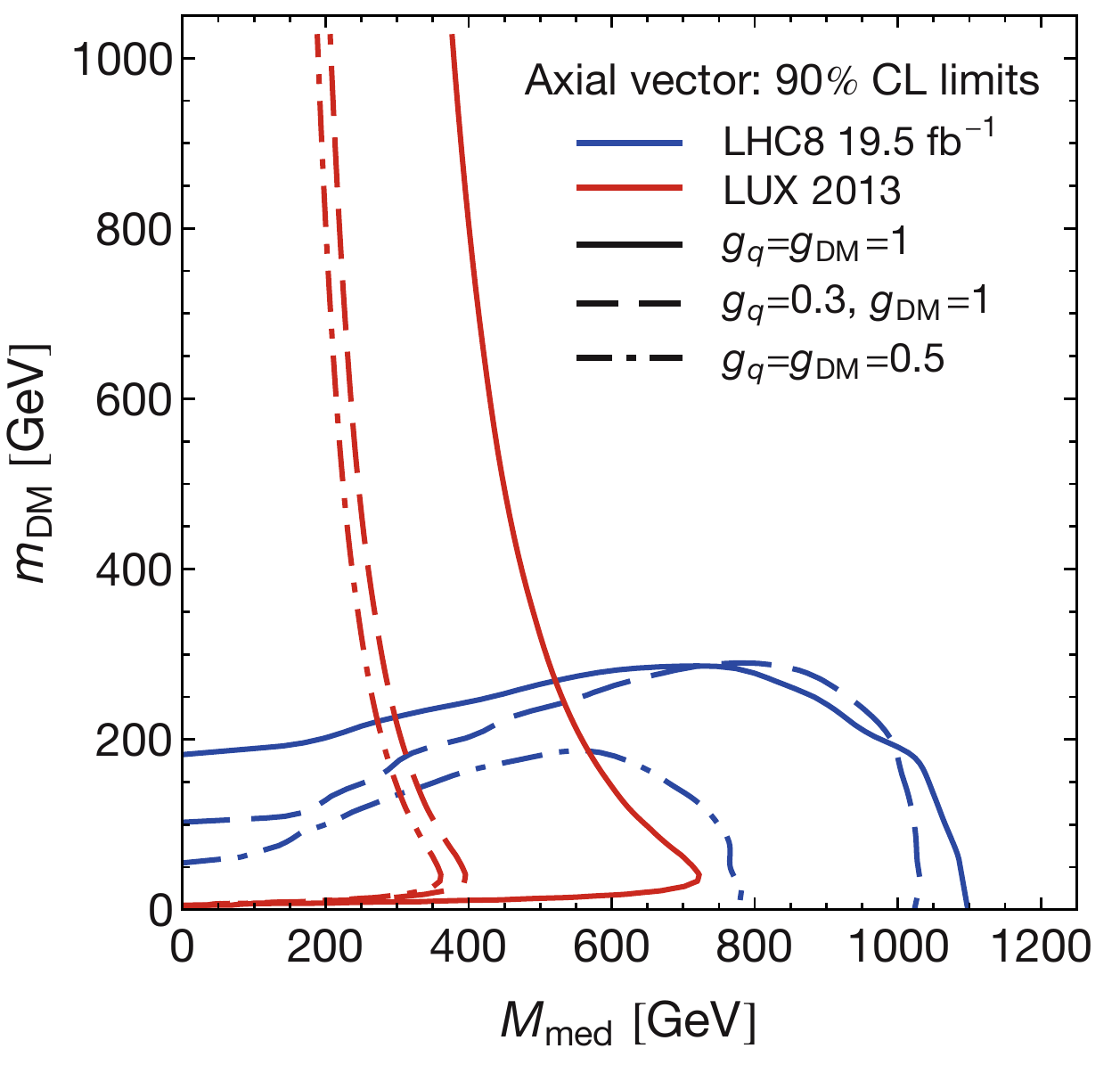}
\caption{The 90\% CL limits from current mono-jet (blue lines) and direct detection (red lines) searches in the $\mDM$ vs~$\mMed$ plane for the vector (left panel) and axial-vector (right panel) mediators. The region to the left of the various curves is excluded. The solid, dashed and dot-dashed lines are for $(\gq,\gDM)=(1,1)$, $(0.3,1)$ and $(0.5,0.5)$ respectively. While the LHC limits are similar in both panels, the LUX limits are significantly more constraining for vector mediators. Note that the vector case has log scales for both axes while the axial-vector case has linear scales.}
\label{fig:Mdm-Mmed}
\end{figure}

We first present our 90\% exclusion limits from the 8~TeV mono-jet and LUX 2013 searches for the MSDM models defined in section~\ref{sec:SMS}. In figure~\ref{fig:Mdm-Mmed} we compare the LHC mono-jet search with the LUX result in the $\mDM$ vs~$\mMed$ plane for the vector (left panel) and axial-vector (right panel) MSDM models. The solid, dashed and dot-dashed lines show three different coupling scenarios:  $(\gq,\gDM)$ = $(1,1)$, $(0.3,1)$ and $(0.5,0.5)$ respectively. The region to the left of the lines is excluded. 

We begin by making some general comments which hold for all of the figures in this section. While the collider limits are similar in sensitivity for the vector and axial-vector cases, the LUX limits on $\mMed$ (for a given value of $\mDM$) for a vector mediator exceed the axial-vector limits by almost two orders of magnitude. The significant difference in the LUX limits is explained by the atomic number squared ($A^2\simeq131^2$) enhancement of the SI cross-section in the case of a vector mediator. The axial-vector case, for which the scattering cross-section is SD, does not exhibit this enhancement (cf.~eqs.~\eqref{eq:SI} and~\eqref{eq:SD}). For the vector mediator, the LUX limits are significantly stronger than the LHC limits. The only exception is at smaller values of the DM mass ($\mDM\lesssim5$~GeV) where direct detection experiments lose sensitivity. We discuss this region further in section~\ref{sec:low}. For axial-vector mediators, the mono-jet and direct detection searches show good complementarity, probing orthogonal directions in the parameter space. While the mono-jet search is more sensitive to large values of $\mMed$, the direct detection experiments extend the reach to large values of $\mDM$. 

For the different coupling scenarios the $\gq=\gDM=1$ example provides the strongest limits for both mono-jet and direct detection searches, while the $g_{q}=\gDM=0.5$ case possesses the weakest sensitivity. The behaviour of the direct detection limits is straightforward to understand: the cross-section scales like $g_q^2 \gDM^2/\mMed^4$ (cf. eqs.~\eqref{eq:SI2MSDM} and~\eqref{eq:SD2MSDM}) and thus, for a given value of $\mDM$, the limit on 
$\mMed$ is proportional to $\sqrt{g_q \gDM}$.

The behaviour of the collider limits is more complex and can be understood as follows. First, consider the collider limits for fixed values of $\mDM$ and at large values of $\mMed$. Here we again expect the cross-section to scale as $g_q^2 \gDM^2$. However, unlike for direct detection, we must also take into account the effect of the mediator width $\Gamma_{\rm{med}}$ as discussed in~\cite{Fox:2011fx,Buchmueller:2013dya}. In this case the partonic cross-section scales approximately as $g_q^2 \gDM^2/(\mMed^4 \Gamma_{\rm{med}})$ so that the limit on $\mMed\propto\sqrt{g_q \gDM}\,\Gamma_{\rm{med}}^{-1/4}$. Although this approximation ignores the PDFs, we find numerically that it gives a good rule of thumb for the scaling at large values of $\mMed$. From eqs.~\eqref{eq:Gamma1} to~\eqref{eq:Gamma4} we see that at large values of $\mMed$ the width of the mediator $\Gamma_{\rm{med}}$ is proportional to $18\gq^2 + \gDM^2$. This implies that the $(\gq,\gDM)=(0.3,1)$ case is enhanced with respect to the other cases because $\Gamma_{\rm{med}}$ is smallest for this case. This enhancement explains why the $(\gq,\gDM)=(0.3,1)$ mono-jet limit is closer to the $\gq=\gDM=1$ limit rather than the $g_{q}=\gDM=0.5$ limit as in the case of the direct detection limits.

Second, consider the collider limits for fixed values of $\mMed$. The limits on $\mDM$ are constrained by the energy of the colliding partons since two DM particles must be produced in the final state. The phase-space suppression factors that enter the cross-section for vector and axial-vector mediators are typically of the form $\sqrt{Q_{\rm{tr}}^2-4\mDM^2}(Q_{\rm{tr}}^2+2\mDM^2)$ and $(Q_{\rm{tr}}^2-4\mDM^2)^{3/2}$ respectively, where $Q_{\rm{tr}}\simeq700$~GeV is the s-channel momentum transfer~\cite{Busoni:2014sya}. It should be noted that these phase space factors also appear in the width calculation cf.~eqs.~\eqref{eq:Gamma1} to~\eqref{eq:Gamma4}. The axial-vector mediator is more strongly phase-space suppressed, which accounts for the greater suppression between the $\gq=\gDM=1$ and $\gq=\gDM=0.5$ limits at small $\mMed$ in the axial-vector case compared to the vector case in figure~\ref{fig:Mdm-Mmed}. Note that these phase-space suppression factors also account for the difference between the vector and axial-vector EFT limits in the left panel of figure~\ref{fig:validation}.

\begin{figure}[t!]
\centering
\includegraphics[width=0.495\columnwidth]{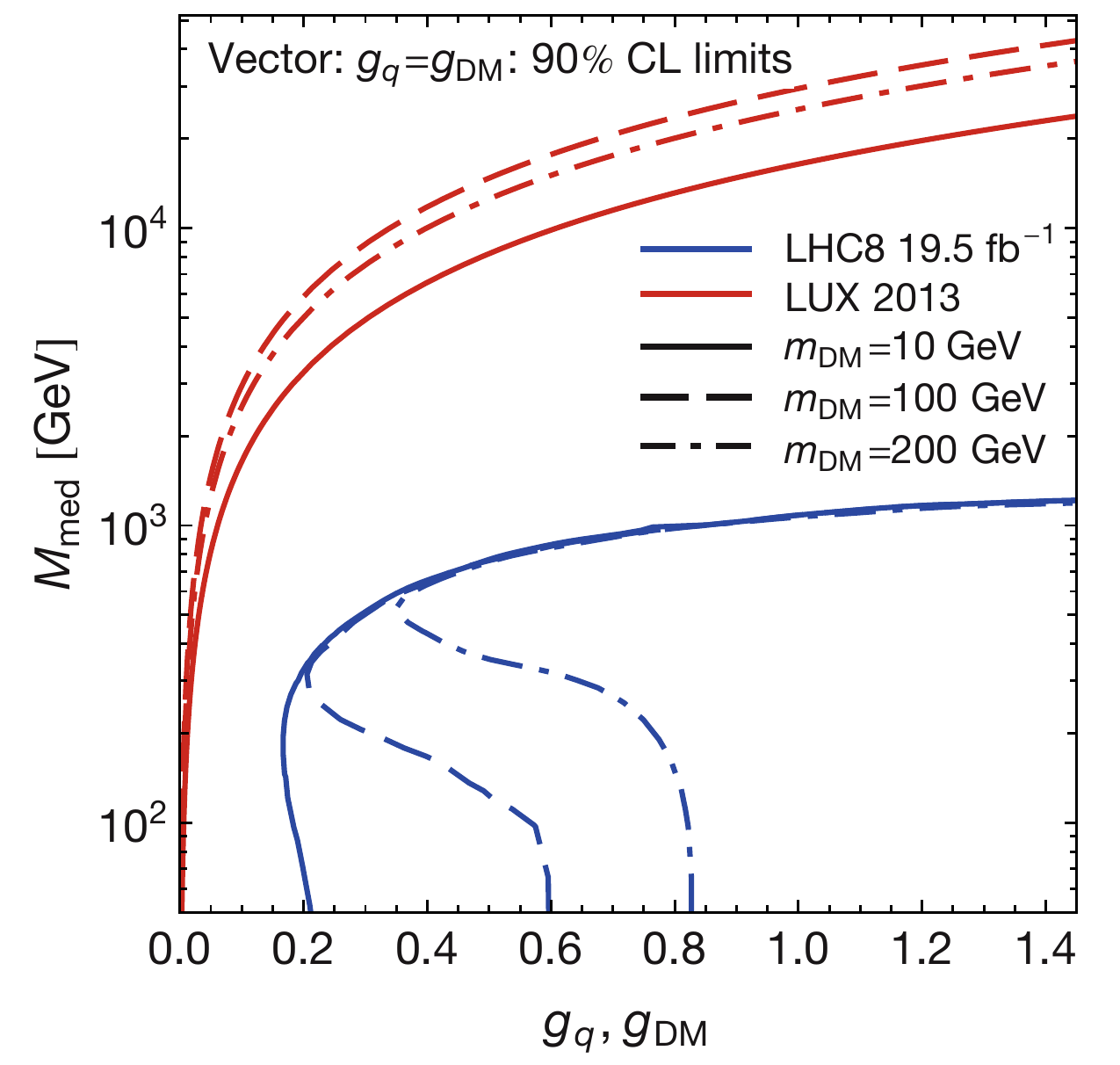}
\includegraphics[width=0.495\columnwidth]{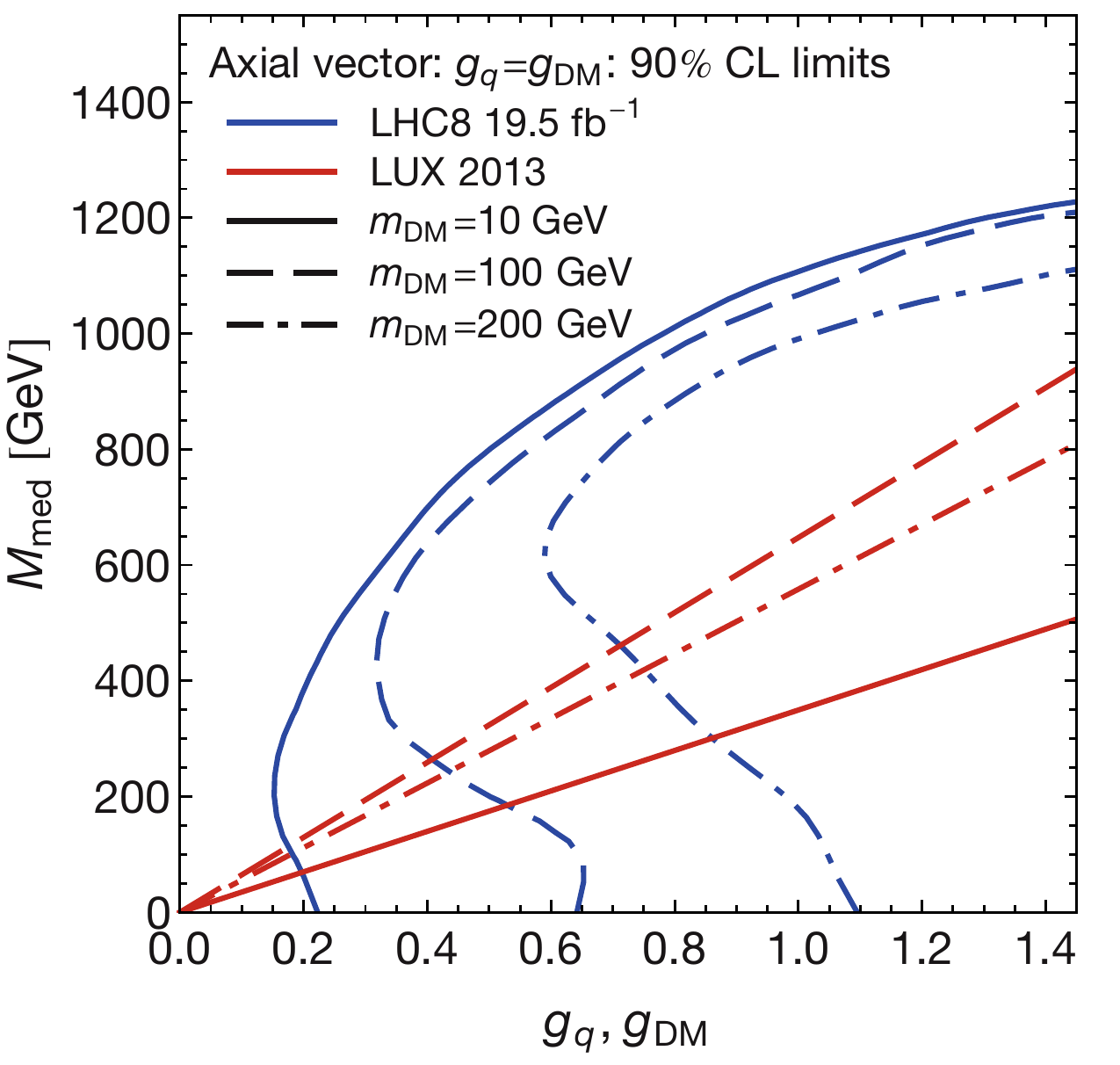}
\caption{The 90\% CL limits from the LHC mono-jet (blue lines) and LUX (red lines) searches in the $\mMed$ vs ($g_{q}=\gDM$) plane; we have fixed $g_q=\gDM$. The left and right panels show the limits for vector and axial-vector mediators respectively. The region to the right of the various curves is excluded. We show three different dark matter masses: $\mDM=10$~GeV is solid, $\mDM=100$~GeV is dashed and $\mDM=200$~GeV is dot-dashed. Note that the $\mMed$-axis has a log (linear) scale in the left (right) panel.}
\label{fig:Mmed-couplings}
\end{figure}

Further insights into the dependence on the chosen coupling scenarios can be gained by looking at the projection in the $\mMed$ vs $\gq,\gDM$ plane, shown in figure~\ref{fig:Mmed-couplings}. The solid, dashed and dot-dashed lines show the limits for $\mDM=10,\,100$ and~$200$~GeV respectively. We have fixed $g_q=\gDM$ in this figure and the region to the right of the lines is excluded.

In this plane the mono-jet limits are similar for axial-vector and vector mediators: both exclude down to $g_q=\gDM\simeq0.2$ for light mediators (when $\mDM=10$~GeV) and both show a characteristic turning point owing to the resonance of the s-channel mediator. The resonance occurs when $\mMed^2\gtrsim 4 \mDM^2+\MET^2$ (where $\MET=400$~GeV). For values of $\mMed$ below this, the limits on the couplings become weaker because the production is through an off-shell mediator. The small difference in behaviour between the vector and axial-vector limits at large couplings and large $\mMed$ can again be understood by the different phase-space suppression factors. The axial-vector phase-space suppression is stronger so there is more of a difference between the limits at $\mDM=10,\,100$ and~$200$~GeV. We again note that this behaviour is also found in the EFT limits in the left panel of figure~\ref{fig:validation} for the same reason.
 
The direct detection limit curves instead show a rather simple behaviour as there are no resonance effects in this case. The limits on $\mMed$ rise in proportion to the coupling strength cf.~eqs.~\eqref{eq:SI2MSDM} and~\eqref{eq:SD2MSDM}. In comparison to the collider limits where the limits are stronger for smaller $\mDM$, the direct detection limit is strongest at $\mDM=100$~GeV and weakest for $\mDM=10$~GeV. This can be easily understood with reference to figure~\ref{fig:DDvalidation}.

\begin{figure}[t!]
\centering
\includegraphics[width=0.495\columnwidth]{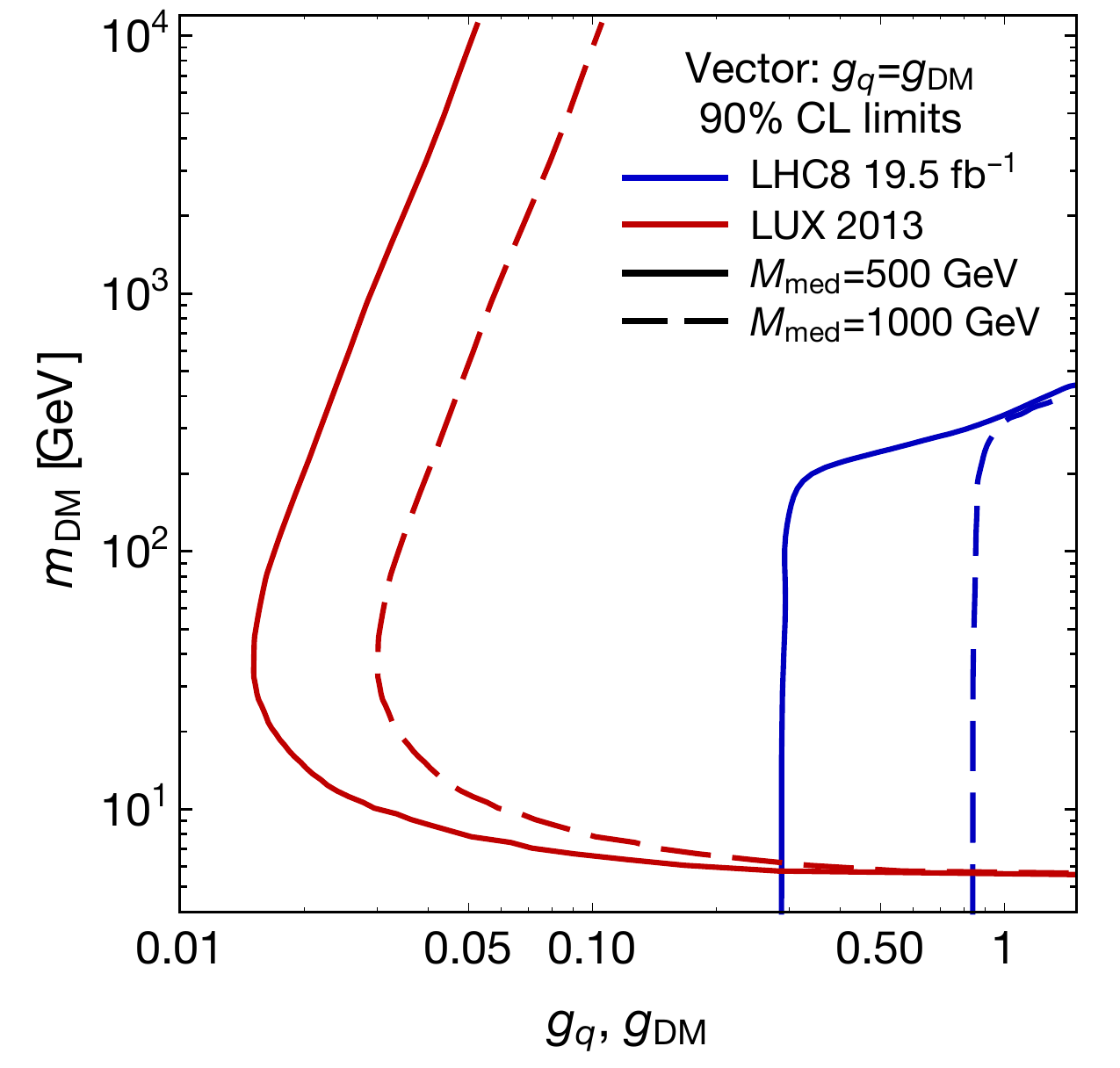}
\includegraphics[width=0.495\columnwidth]{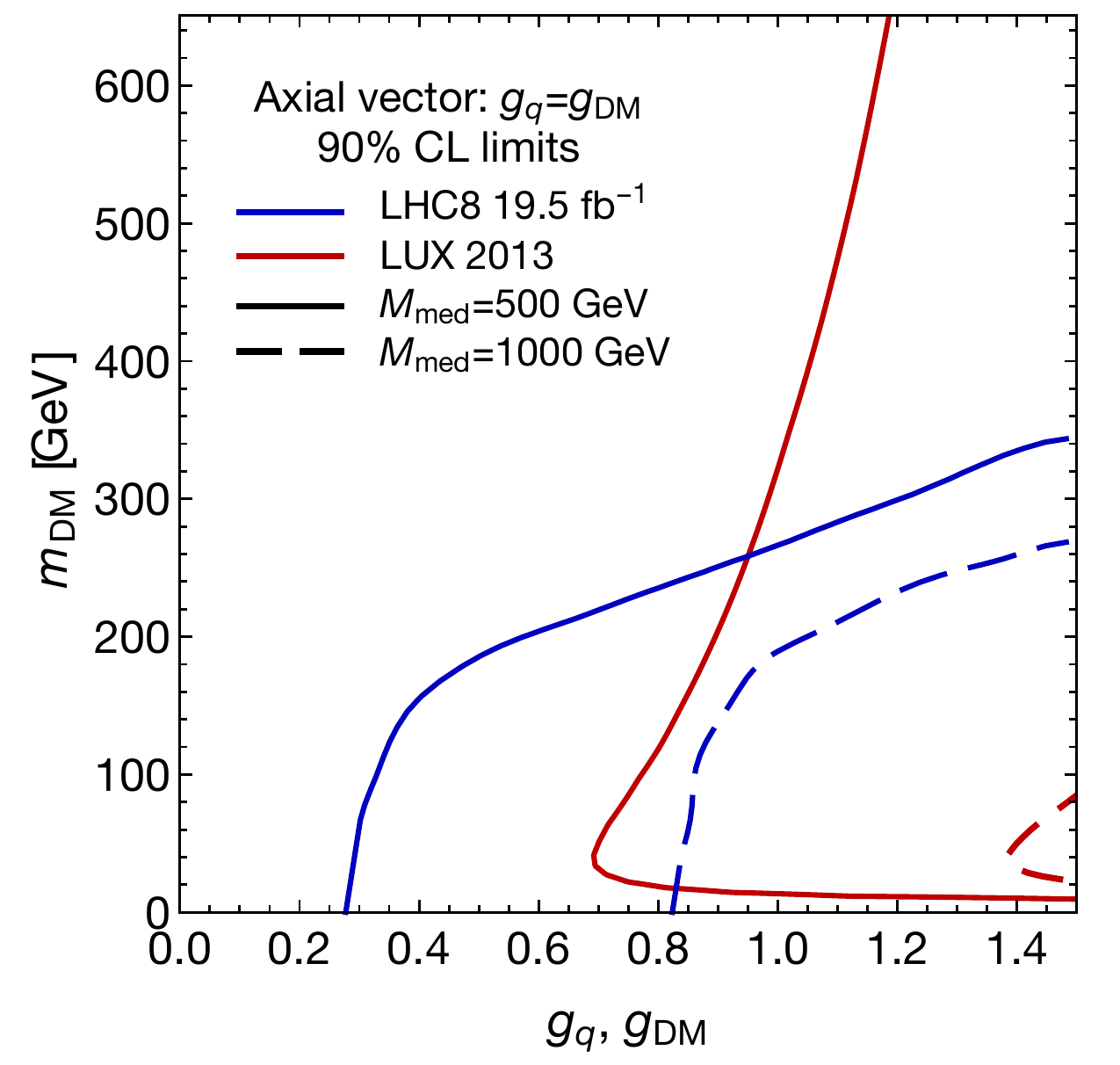}
\caption{The 90\% CL limits from mono-jet (blue lines) and direct detection (red lines) searches in the $\mDM$ vs ($g_{q}=\gDM$) plane for the vector (left) and axial-vector (right) mediators. We have fixed $g_q=\gDM$. The parameter space to the right of the various curves is excluded. We show two different mediator masses: $\mMed=500$~GeV is solid and $\mMed=1000$~GeV is dashed. Note that the $\mDM$-axis scales are different for each panel.}
\label{fig:Mdm-couplings}
\end{figure}

Figure~\ref{fig:Mmed-couplings} again demonstrates the good complementarity between the mono-jet and direct detection searches for axial-vector mediators since they probe different regions in the MSDM parameter space. The direct detection experiments are better at probing small $\mMed$, which are not yet accessible to the collider searches as the mediator is very off-shell, while the LHC is a better probe at larger values of~$\mMed$.

In figure~\ref{fig:Mdm-couplings} we compare the two searches in the $\mDM$ vs  ($\gq=\gDM$) plane. The solid and dashed lines show the limits for $\mMed=500$ and~$1000$~GeV respectively. The region to the right of the curves is excluded in both panels. The behaviour of the limits in this figure is similar to that shown already. The LUX limit is significantly stronger than the LHC limit for vector mediators, except in the low $\mDM$ region. The LUX and LHC limits show good complementarity in the axial-vector case as they probe different regions of parameter space. The scalings of the collider limits can be understood with reference to the width and phase-space scalings discussed in connection with figure~\ref{fig:Mdm-Mmed}. As discussed previously, for a given value of $\mDM$ the LUX limit on $\sqrt{g_q \gDM}\propto\mMed$.

\begin{figure}[t!]
\centering
\includegraphics[width=0.495\columnwidth]{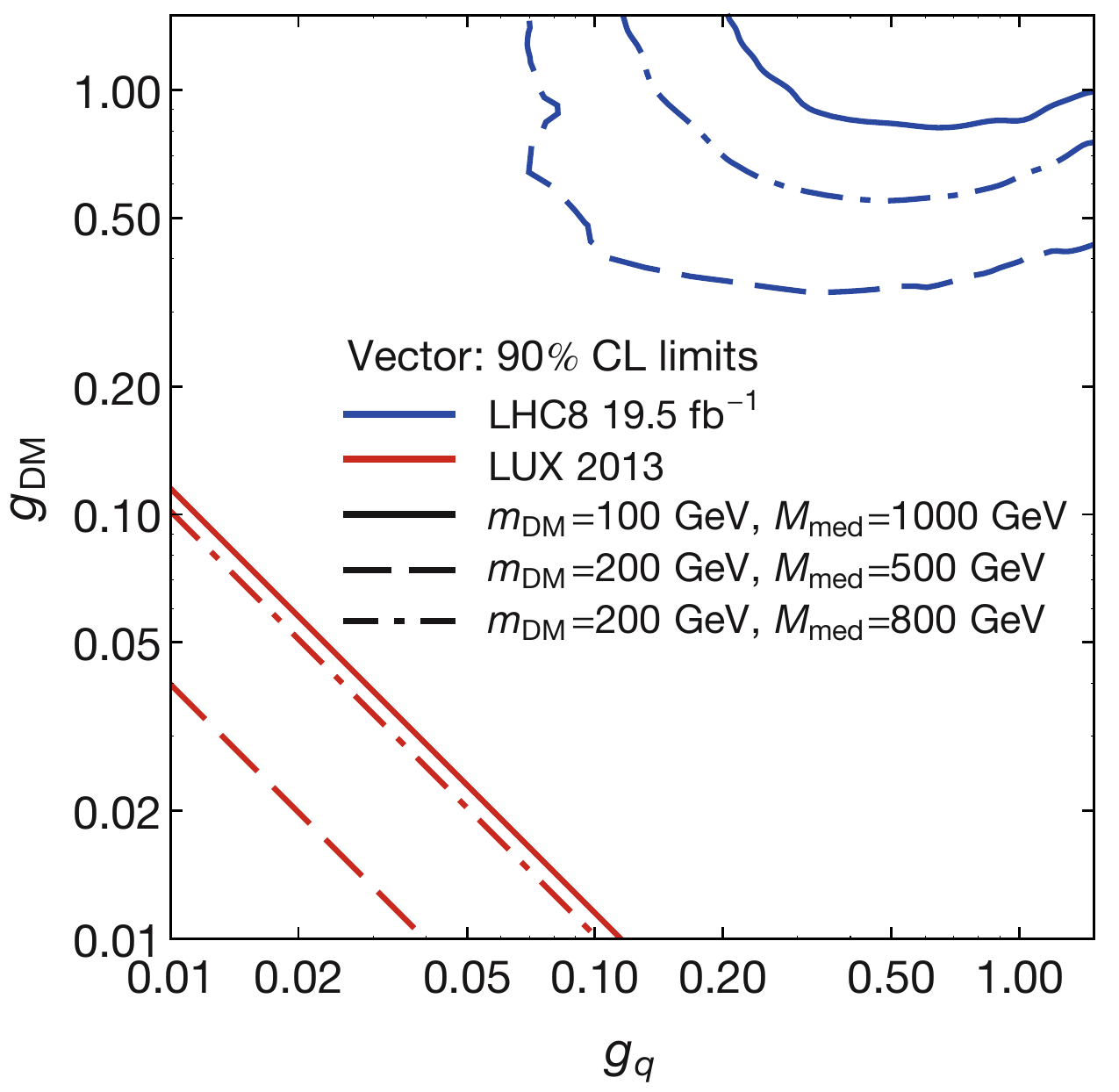}
\includegraphics[width=0.495\columnwidth]{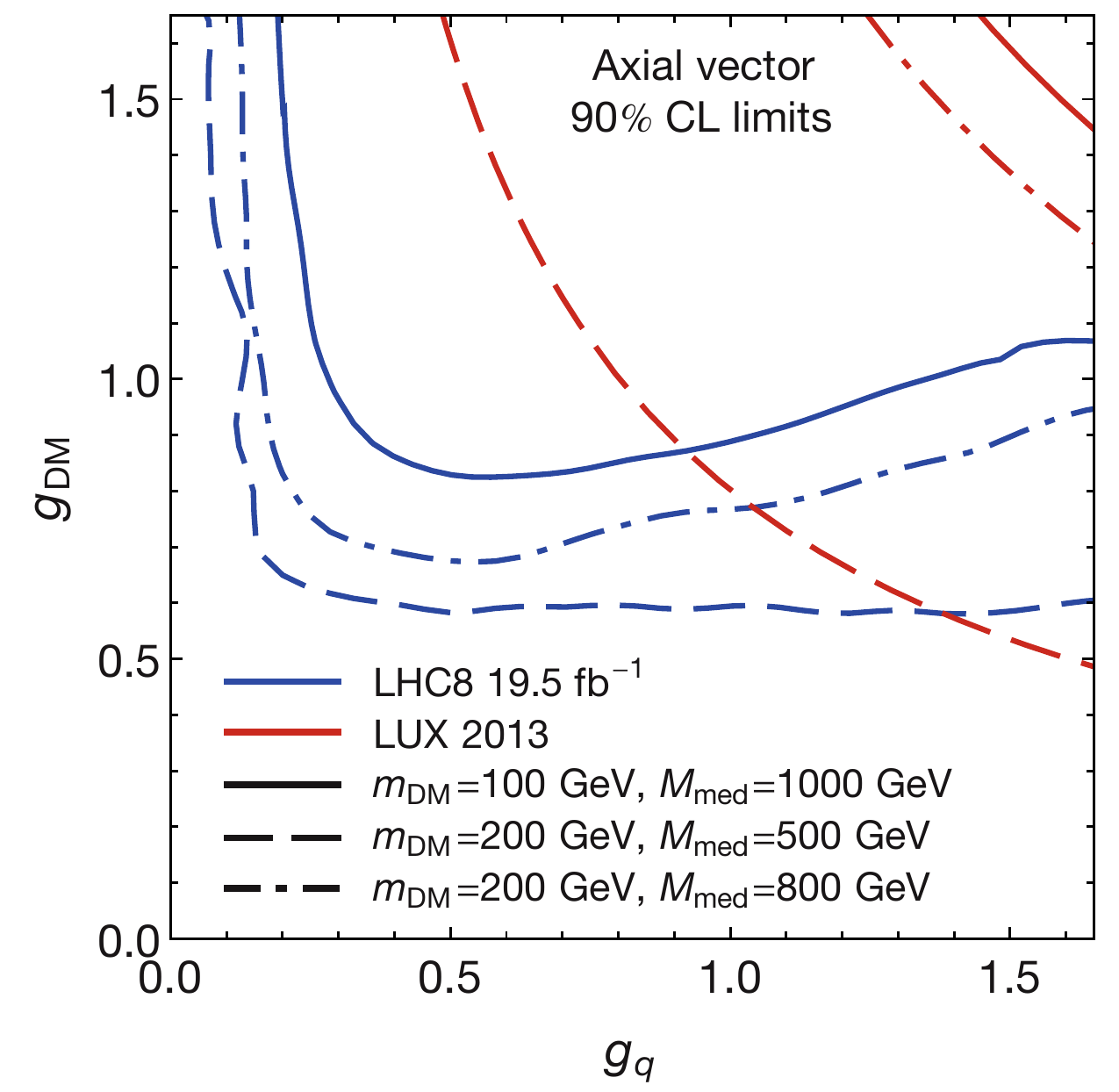}
\caption{The current LHC mono-jet (blue lines) and LUX (red lines) 90\% CL limits in the $\gDM$ vs $g_q$ plane for a vector (left panel) and axial-vector (right panel) mediator. The parameter space above and to the right of the various lines is excluded. We show three different sets of dark matter and mediator masses $(\mDM,\mMed)$: $(100,1000)$~GeV is solid, $(200,500)$~GeV is dashed and $(200,800)$~GeV is dot-dashed. Note that the left (right) panel has log (linear) axes.}
\label{fig:couplings}
\end{figure}
\sloppy
Finally, we consider the limits in the $\gDM$ vs $g_q$ plane, where we fix both $\mMed$ and $\mDM$. Figure~\ref{fig:couplings} shows that the direct detection limits are fully symmetric in this plane. This is because the direct detection cross-section is sensitive only to the product $\gq^2 \gDM^2$ cf.~eqs.~\eqref{eq:SI2MSDM} and~\eqref{eq:SD2MSDM}. However, the mono-jet search is able to break this degeneracy because it is also sensitive to the mediator width, which is not symmetric in $\gq$ and $\gDM$ ($\Gamma_{\rm{med}}\propto18\gq^2 + \gDM^2$ at large values of $\mMed$). Therefore in the event of a DM discovery at colliders and direct detection, the mono-jet analysis, or other collider searches like the dijet or jets plus~MET searches, could add important information in order to disentangle the coupling structure.
\fussy

\subsection{Low dark matter mass region}
\label{sec:low}

We now focus on the low $\mDM$ region of the vector mediator parameter space. This is of particular interest as direct detection searches lose sensitivity for $\mDM\lesssim 10$~GeV because the momentum transfer becomes small and the nuclear recoil energy falls below experimental thresholds. This is not an issue for collider searches and so it is interesting to understand how collider searches can help to constrain this parameter space.  It is also an interesting region both from a theoretical perspective, since $\mDM\simeq 5$~GeV is predicted in many models of asymmetric DM (see~\cite{Petraki:2013wwa} for a recent review), and from a phenomenological perspective, since it is the region where CoGeNT~\cite{Aalseth:2010vx,Aalseth:2011wp,Aalseth:2012if,Aalseth:2014eft},  CRESST-II~\cite{Angloher:2011uu}, CDMS-Si~\cite{Agnese:2013rvf} and DAMA/LIBRA~\cite{Bernabei:2013xsa} reported signal-like excesses in recent years. However, in 2013 both LUX and SuperCDMS reported results which naively exclude these signals. See also~\cite{Davis:2014bla},~\cite{Brown:2011dp,Kuzniak:2012zm,Angloher:2014dua} and~\cite{Blum:2011jf,Pradler:2012qt,Davis:2014cja} for additional non-DM explanations of the CoGeNT, CRESST-II and DAMA/LIBRA excesses. In this section we complement the LUX result with the recent result from SuperCDMS as it extends the sensitivity of direct detection experiments to lower values of $\mDM$. Further details about the SuperCDMS result and how it is used are provided in section~\ref{sec:DD}.

\begin{figure}[t!]
\centering
\includegraphics[width=0.495\columnwidth]{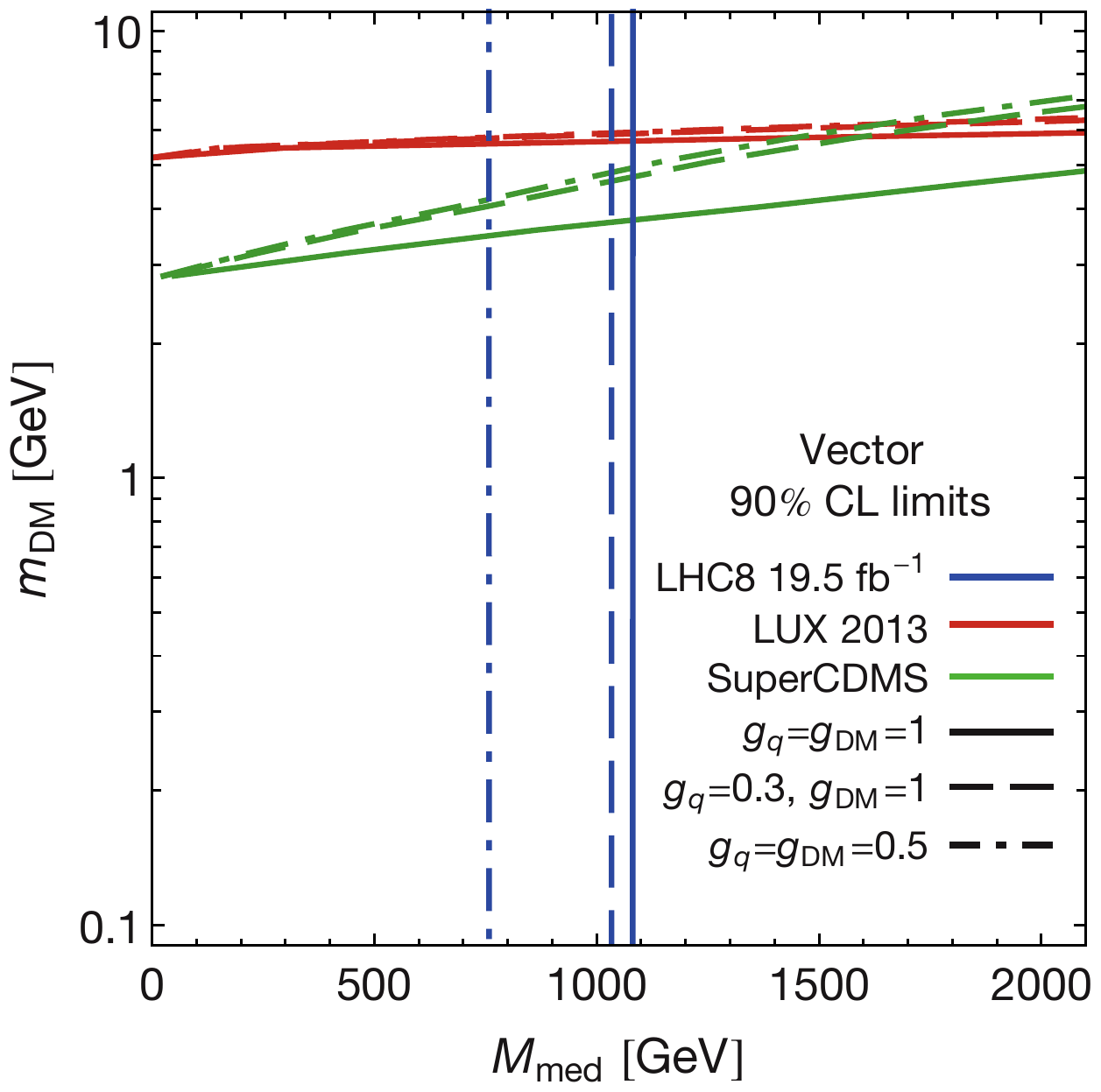}
\includegraphics[width=0.495\columnwidth]{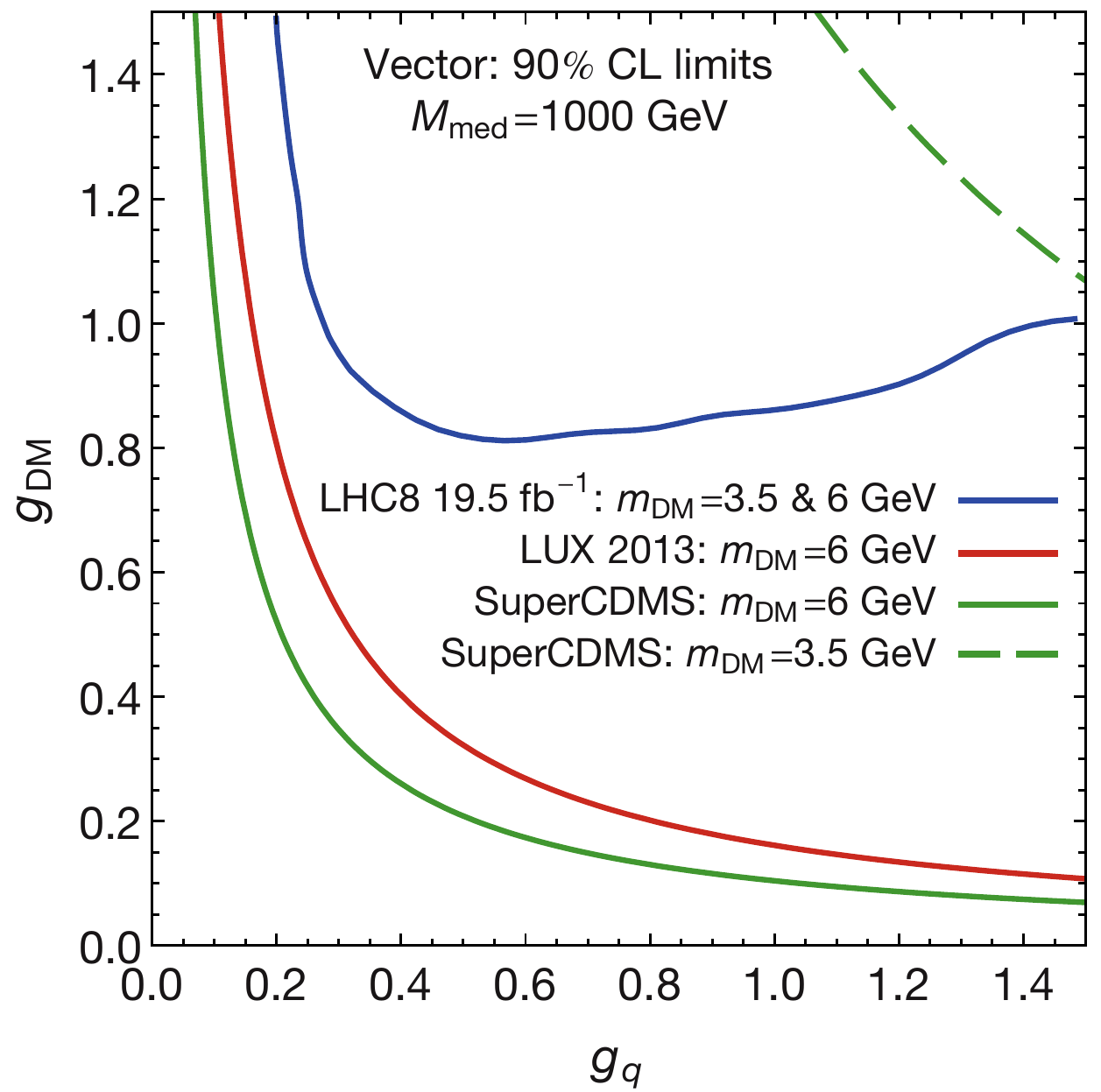}
\caption{The 90\% CL limits from the LHC mono-jet (blue lines), LUX (red lines) and SuperCDMS (green lines) searches in the low $\mDM$ region. The left and right panels show the $\mDM$ vs $\mMed$ and $\gDM$ vs $g_q$ planes for a vector mediator. In the left panel, the region to the left and above the lines is excluded. In the right panel, the region to the right and above the lines is excluded. In the $\mMed$-$\mDM$ plane we show three different set of couplings: $(\gq,\gDM)=(1,1)$ is solid, $(0.3,1)$ is dashed and $(0.5,0.5)$ is dot-dashed. In the $\gq$-$\gDM$ plane we fix $\mMed=1$~TeV and show two different choices of the dark matter mass: $\mDM=3.5$~GeV and $\mDM=6$~GeV. In the right panel, the mono-jet limit is identical for the two different $\mDM$ values while there is no LUX limit for $\mDM=3.5$~GeV.}
\label{fig:low-mass}
\end{figure}

In figure~\ref{fig:low-mass} we show the limits from the LHC mono-jet, SuperCDMS and LUX searches in the $\mDM$ vs $\mMed$ plane (left panel) and the $\gDM$ vs $g_q$ plane (right panel). In the left panel we show again the three different coupling scenarios: $(\gq,\gDM)$= $(1,1)$, $(0.3,1)$ and $(0.5,0.5)$. SuperCDMS and LUX exclude the region above the green and red lines, while the LHC limits exclude the region to the left of the blue lines. In the right panel we show the limits for $\mDM=3.5$ and~$6$~GeV. We fix $\mMed=1$~TeV as this approximates the current sensitivity of the mono-jet searches (see figure~\ref{fig:Mdm-Mmed}) but a lower mediator mass would not significantly change our discussion. The region to the right of the various lines is excluded.
  
The left panel demonstrates that in the region of interest the LHC limits are independent of $\mDM$. The LHC bounds are only limited by on-shell mediator production and currently extend to $\mMed\simeq1100$~GeV (for $\gq=\gDM=1$). This is in contrast to LUX and SuperCDMS, whose sensitivity drops off rapidly below 6~GeV and 3.5~GeV respectively. This is also demonstrated in the right panel. While the LHC limit is independent of the choices for $\mDM$, the LUX and SuperCDMS limits drop off rapidly. These examples demonstrate that direct detection and collider limits have good complementarity for vector mediators in the low $\mDM$ region. This mass region is highly motivated in asymmetric DM models, where typically $\mDM\simeq 5$~GeV, so mono-jet searches may have an important role to play in testing these models (see e.g.~\cite{MarchRussell:2012hi}).

\subsection{Projection for future searches}
\label{sec:pro}

In this section we provide extrapolations of how the limits and complementarity between the LHC and direct detection search avenues will continue to develop. Both the collider and direct detection communities have plans for mid- and long-term projects that possess the potential to significantly increase the sensitivity for DM searches.  
\begin{figure}[t!]
\centering
\includegraphics[width=0.495\columnwidth]{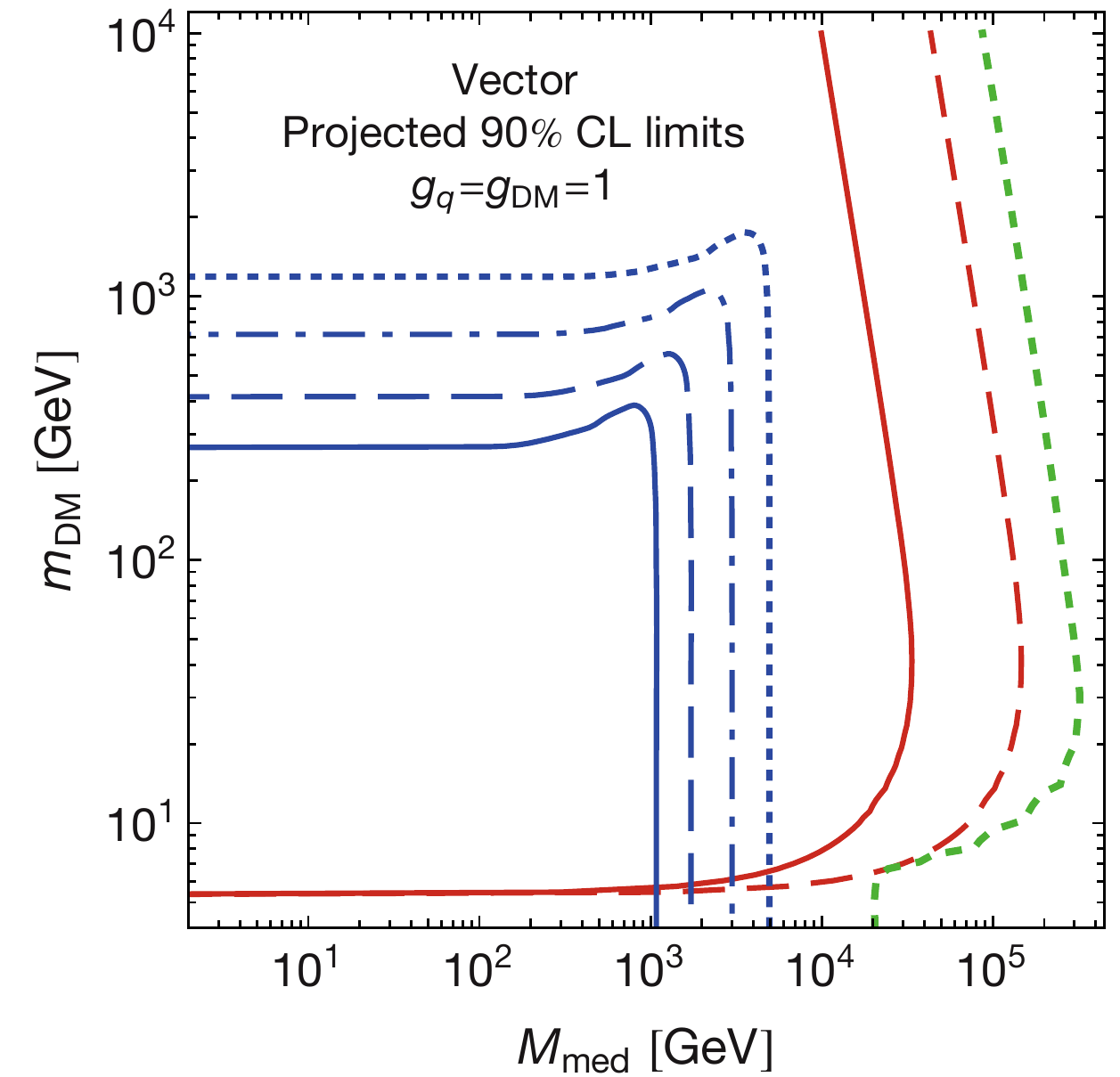}
\includegraphics[width=0.495\columnwidth]{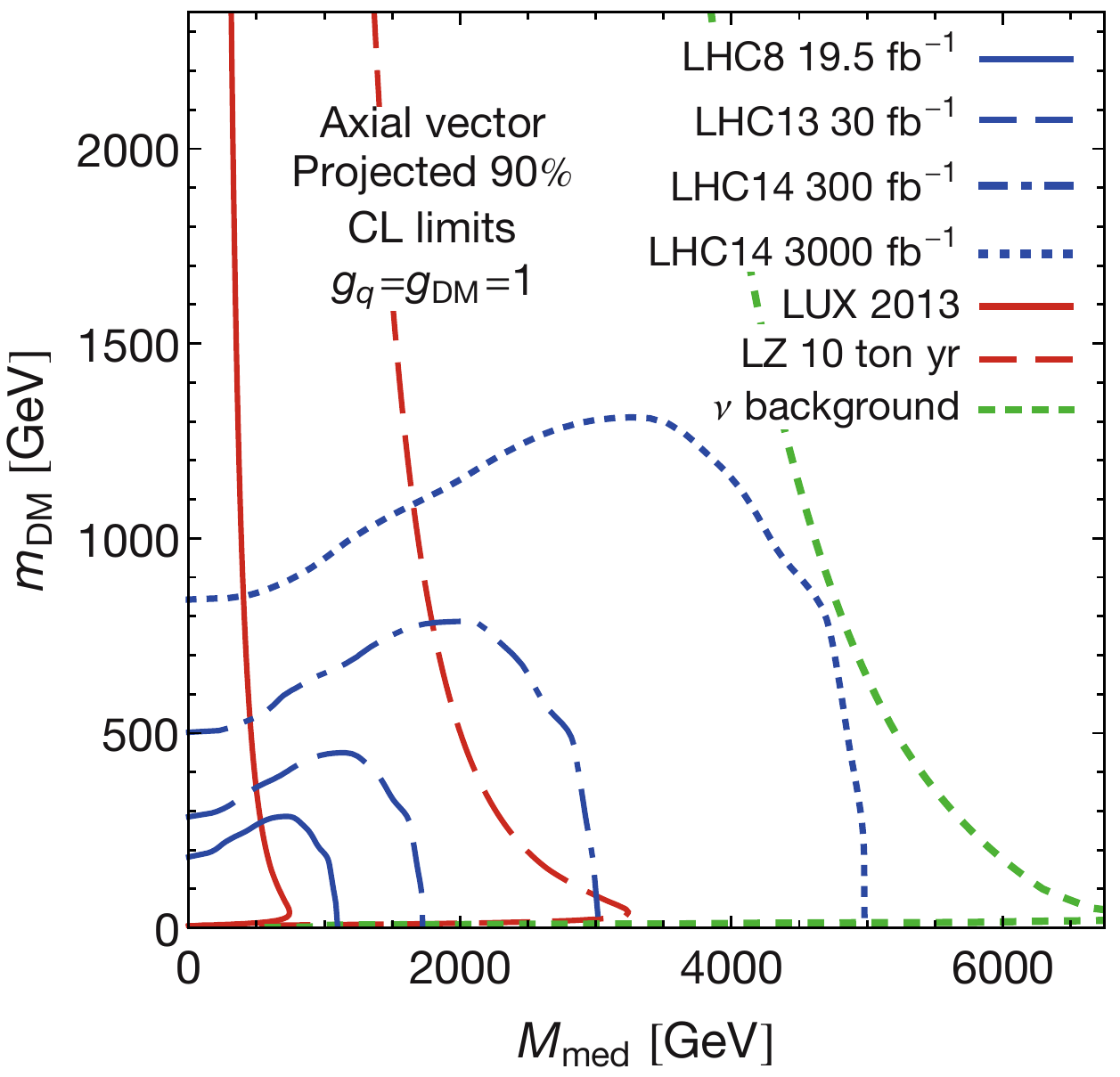}
\caption{The blue and red lines show the current and projected 90\% CL limits from the LHC mono-jet and LUX searches in the $\mDM$ vs $\mMed$ plane. The left and right panels show the limits for vector and axial-vector mediators respectively for $(\gq,\gDM)=(1,1)$. The region to the left of the various curves is excluded. The plot legend is the same for both panels. The short-dashed green lines shows the direct detection discovery reach after accounting for the neutrino background. While LUX has better sensitivity than mono-jet searches and approaches the neutrino limit for vector mediators, the opposite is true for axial-vector mediators. Note that the left (right) panel has log (linear) axes.}
\label{fig:proj:Mdm-couplings}
\end{figure}

For the LHC we provide projected limits for:
\begin{itemize}
\item LHC 13 TeV and 30 $\mathrm{fb}^{-1}$. This gauges the reach for the first year of LHC running in 2015. 
\item LHC 14 TeV and 300 $\mathrm{fb}^{-1}$. This provides an estimate of the ultimate reach of the LHC.
\item HL-LHC 14 TeV and 3000 $\mathrm{fb}^{-1}$. This is the expected reach of a high-luminosity upgrade of the LHC. 
\end{itemize}
The basis for these extrapolations are the 8 TeV limits of the CMS mono-jet search presented in section~\ref{sec:results}. These limits are scaled to the different future scenarios assuming that the underlying performances of the search in terms of signal efficiency and background suppression remains unchanged.
These assumptions were also used in the Snowmass~\cite{CMS:2013xfa} and ECFA~\cite{CMS-PAS-FTR-13-014,ATL-PHYS-PUB-2014-010,ATL-PHYS-PUB-2014-007} studies and form the basis of the Collider Reach~\cite{colliderreach} tool. Ref.~\cite{colliderreach} also shows the good agreement between using this extrapolation and using a full simulation. Furthermore, the underlying assumption of maintaining the present performance of the searches is supported by the ATLAS and CMS upgrade programmes,  which both put forward this assumption as the main upgrade goal.   

For the direct detection experiments we show two different scenarios:
\begin{itemize}
\item LZ with 10 ton year exposure. This is our estimated limit for the lifetime exposure of the LZ experiment~\cite{Malling:2011va}. The successor to XENON1T~\cite{Aprile:2012zx} should achieve a similar sensitivity.
\item The xenon discovery reach when accounting for the coherent neutrino scattering background~\cite{Billard:2013qya}. 
\end{itemize}
These limits are based on the calculations described in section~\ref{sec:valid}.

Figure~\ref{fig:proj:Mdm-couplings} shows our estimates of future limits on the vector mediator (left panel) and axial-vector (right panel) mediators for $(\gq,\gDM)=(1,1)$. We see that the collider limits improve for each of the scenarios we consider. For the first year of operation of the LHC in 2015, we expect the reach of the mono-jet search to go up to $\mMed\simeq1.8$~TeV for $\mDM\approx350$~GeV, while the ultimate reach of the LHC is expected to be around 3~TeV for $\mDM$ up to 750~GeV. For the HL-LHC the mono-jet limit are projected to extend out to 5~TeV for $\mDM$ up to 1~TeV. However, except at very low $\mDM$ where the collider limit is stronger, the LZ limit will be stronger than even the high-luminosity LHC limit for a vector mediator. The LZ sensitivity even approaches the ultimate sensitivity of direct detection experiments determined by the neutrino background.

For the axial-vector mediator, the mono-jet reach is similar to the vector case. However, in this case the mono-jet reach extends beyond the LZ limit for $\mDM\lesssim1$~TeV and nearly extends to the xenon discovery reach from the neutrino background for the choice of couplings we have made here: $(\gq,\gDM)=(1,1)$. For larger couplings the collider can even probe parameter space beyond the xenon discovery reach. However, LZ will be sensitive to mediator masses of 2~TeV for $\mDM$ up to 2~TeV, thus complementing the collider searches by probing the large $\mDM$ parameter space, which in turn will increase the discovery potential.

\section{Comparing EFT and MSDM limits}
\label{sec:compEFTMSDM}

Thus far we have presented results only in our MSDM models. In this section we contrast the results from these simplified models with those derived from the limit on the EFT suppression scale $\Lambda$. When valid, the advantage of the EFT approach is that the limits are model independent, where the suppression scale is simply related to the couplings and mediator mass of the full theory. We have argued in our previous paper~\cite{Buchmueller:2013dya} that the EFT limits for collider searches should be taken \textit{cum grano salis} as in practice, they apply only to a limited set of theories. However, as they are widely used in the theoretical community it is important to explore the failings of the EFT framework in more detail. In the following, we directly compare our MSDM limits with those derived from the model-independent EFT approach. This quantitatively demonstrates that a naive application of the EFT limits can lead to incorrect conclusions about the sensitivity of mono-jet searches and is an example of a well-defined model for which the model-independent EFT limits poorly approximate the underlying limits in a more-complete theory. 

\begin{figure}[t!]
\centering
\includegraphics[width=0.495\columnwidth]{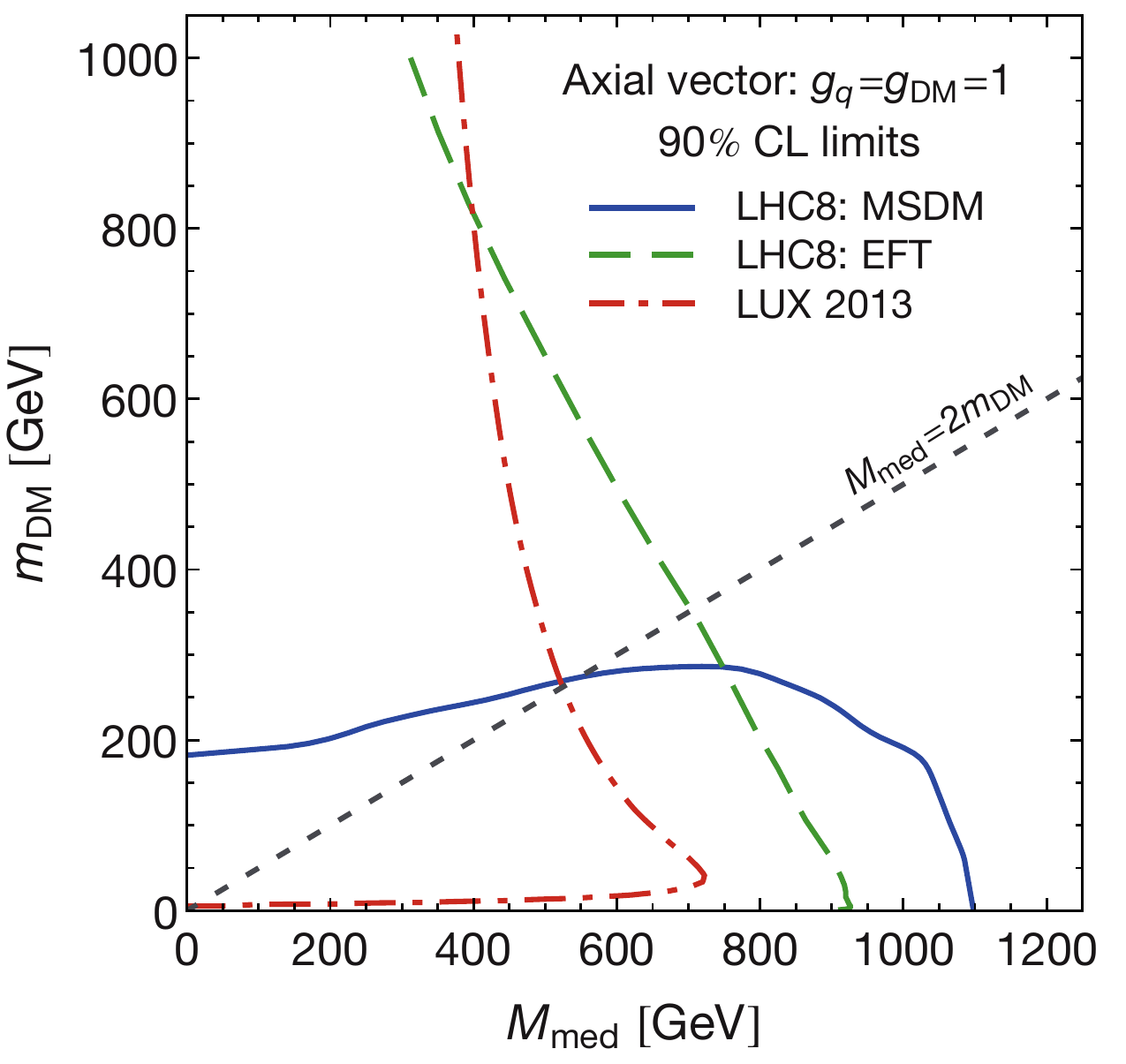}
\includegraphics[width=0.495\columnwidth]{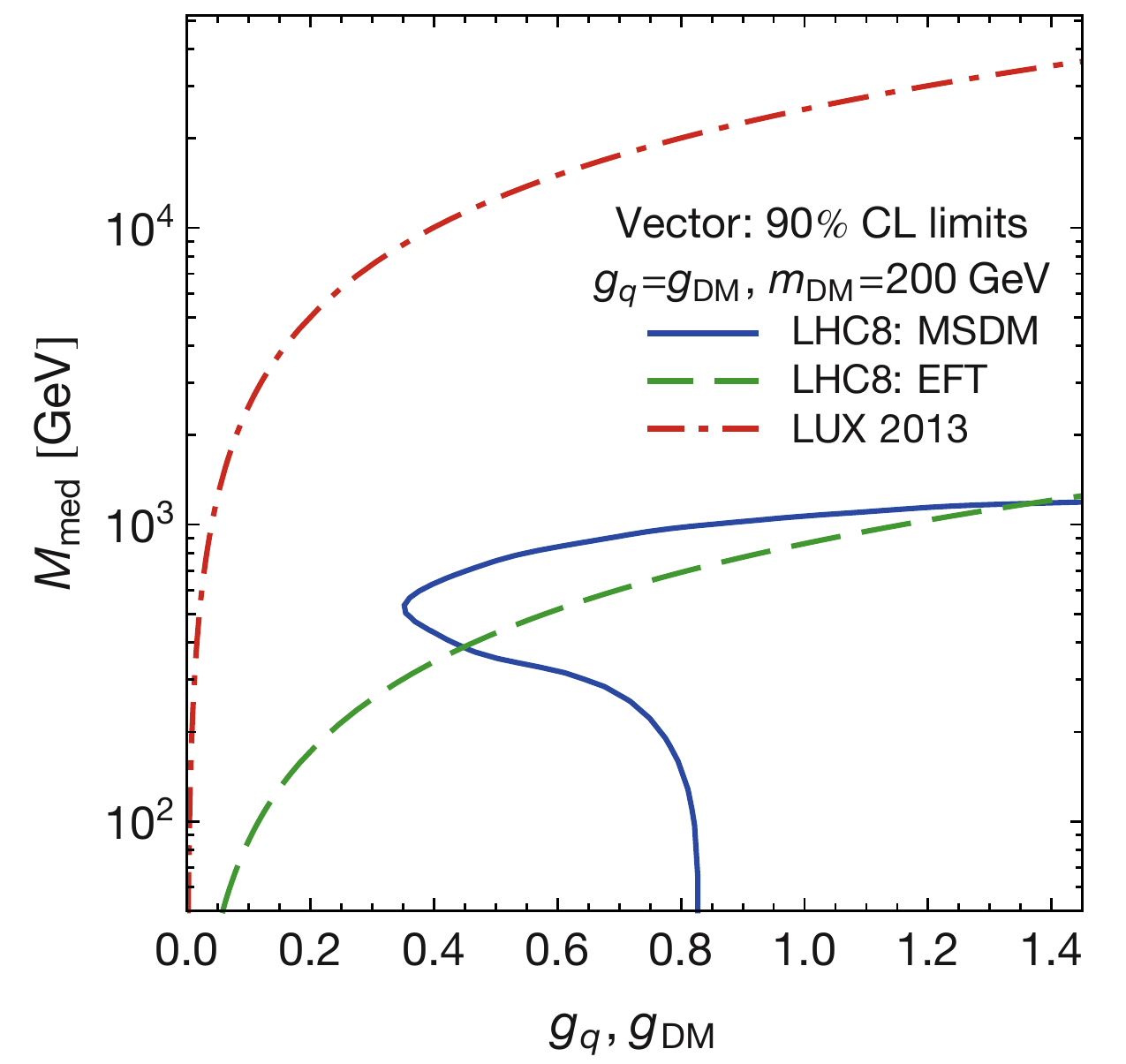}
\includegraphics[width=0.495\columnwidth]{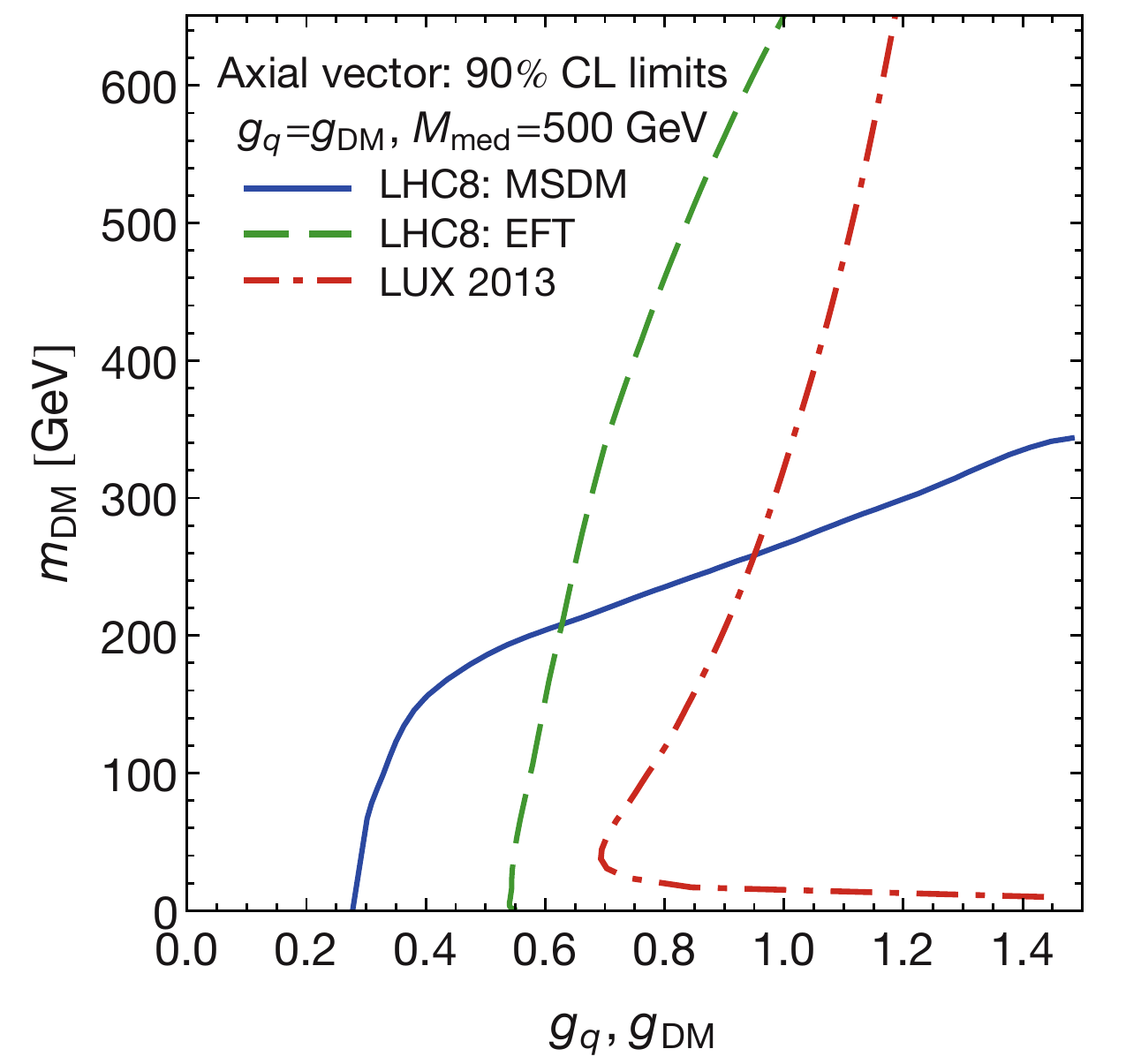}
\includegraphics[width=0.495\columnwidth]{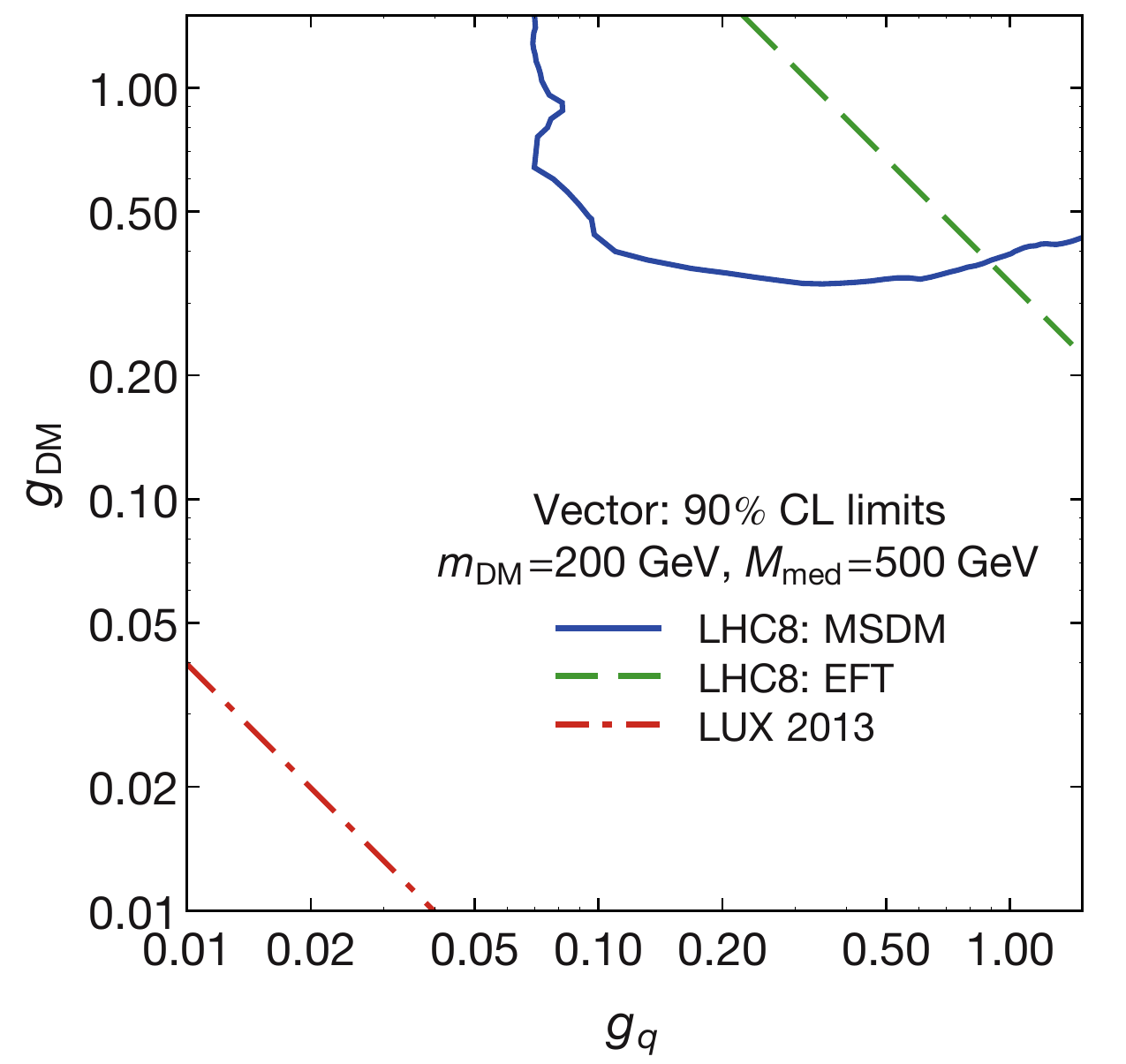}
\caption{Comparisons between the 90\% CL mono-jet limit in our MSDM models (blue solid line) and the EFT framework (green dashed) in the two-dimensional planes considered previously. The red dot-dashed line shows the LUX limit. The left and right panels are for axial-vector and vector mediators respectively.
The MSDM and EFT limits should agree in the domain where the EFT framework is valid. The EFT limits both underestimate the MSDM limit (by missing the resonant enhancement) or overestimate it (by missing off-shell production of the mediator). This may lead to a misleading conclusion regarding the relative sensitivity of mono-jet and direct detection searches. A simple criterion for the validity of the EFT approach is that $\mMed > 2\mDM$. The line $\mMed =2\mDM$ is shown in the upper left panel. Even in the valid region, the EFT limit fails to accurately reproduce the MSDM limit for these parameters.
} 
\label{fig:EFTcomparison}
\end{figure}

When the mediator mass is sufficiently heavy to be safely integrated out, the effective higher dimension operators from our vector and axial-vector MSDM models are
\begin{align}
\mathcal{L}^{\mathrm{eff}}_{\mathrm{vector}}&\supset\sum_q \frac{g_q \gDM}{\mMed^2} \bar{q}\gamma_{\mu}q\,\bar{\chi}\gamma^{\mu}\chi\\
\mathcal{L}^\mathrm{eff}_{\rm{axial}}&\supset\sum_q\frac{g_q \gDM}{\mMed^2} \bar{q}\gamma_{\mu}\gamma^5q\,\bar{\chi}\gamma^{\mu}\gamma^5\chi\;,
\end{align}
where the sum is over all quarks. Comparing with eqs.~\eqref{eq:EFT:vec} and~\eqref{eq:EFT:axvec}, we observe that the relation $\Lambda=\mMed/\sqrt{\gq \gDM}$ holds. Therefore, the EFT limits can be applied to the MSDM models by using the relation $\Lambda=\mMed/\sqrt{\gq \gDM}$ and the CMS 90\%~CL limits on~$\Lambda$, which are shown in the left panel of figure~\ref{fig:validation} as a function of~$\mDM$. In figure~\ref{fig:EFTcomparison} we show a comparison of the current MSDM mono-jet limit (blue solid line) with the naive limit obtained in the EFT framework (green dashed line) for each of the four parameter planes shown in figures~\ref{fig:Mdm-Mmed} to~\ref{fig:couplings}. The EFT limit is naive because we assume that it applies to the full parameter space of the MSDM model. For comparison, we also include the LUX 2013 limit as the red dot-dashed line. The left (right) panels show the limits for an axial-vector (vector) mediator. We again note that the EFT and MSDM provide identical results for the direct detection experiments in this paper, as long as $\mMed \gtrsim100$~MeV.

The first general observation that we can make is that the EFT limits consistently give a poor approximation to the underlying limits obtained in the MSDM model. Not only do the EFT limits exclude different parameter values but also the shape of the limit curve differs dramatically from those of the MSDM limits. This is because the EFT framework does not account for the mediator propagator, and thus does not include the effects of resonant enhancement or off-shell production of the mediator.

The top-left panel of figure~\ref{fig:EFTcomparison} shows the limits in the $\mDM$ vs $\mMed$ plane for an axial-vector mediator when $\gq=\gDM=1$. We observe that at low $\mDM$ the EFT limit is too weak. This is because the EFT framework fails to take into account the resonant enhancement from on-shell mediator production. At larger $\mDM$ the mediator is off-shell and the EFT framework dramatically overstates the limit. No limit is obtained in the MSDM model for $\mDM\gtrsim300$~GeV while a naive application of the EFT limit gives the false impression that the limit extends beyond $\mDM=1$~TeV. This overstating of the limit at high values of $\mDM$ has sometimes led to the wrong conclusion that for spin-dependent interactions, the mono-jet searches outperform direct detection searches. However as the right panels of figures~\ref{fig:Mdm-Mmed} to~\ref{fig:couplings} demonstrate, the two searches probe different orthogonal and complementary regions of the axial-vector parameter space. 

The bottom-left panel shows the limits in the $\mDM$ vs ($g_{q}=\gDM$) plane for an axial-vector mediator with $\mMed=500$~GeV. In this plane, it is particularly clear that naively applying the EFT limit obliterates the complementarity between the direct detection and mono-jet results. The MSDM model reveals that the collider limit on $\gq\,,\gDM$ is stronger at smaller $\mDM$ while the higher $\mDM$ limit is kinematically suppressed because of off-shell mediator production.

The bottom-right panel shows the limits for a vector mediator when $\mMed=500$~GeV and $\mDM=200$~GeV in the $\gDM$ vs $g_q$ plane. We see again how the EFT framework misses important physical effects: the EFT limit is symmetric in $\gq$ and $\gDM$ while the MSDM limit is asymmetric because the mediator width breaks the degeneracy between $\gq$ and $\gDM$. Therefore, the collider possesses sensitivity to the underlying coupling structure, which is not resolved in the EFT approach. 

The final panel in this figure is the top-right panel, which shows the limits for a vector mediator when $\mDM=200$~GeV in the $\mMed$ vs ($g_{q}=\gDM$) plane. We see that the EFT limit again overstates the limit at low $\mMed$ as the EFT framework does not account for the off-shell mediator production. This panel is also the only case where the EFT limit asymptotes to the MSDM limit. This occurs at large couplings and large mediator masses where $\Gamma_{\rm{med}}\gtrsim\mMed$, as discussed in our previous paper~\cite{Buchmueller:2013dya}.

\begin{figure}[t!]
\centering
\includegraphics[width=0.495\columnwidth]{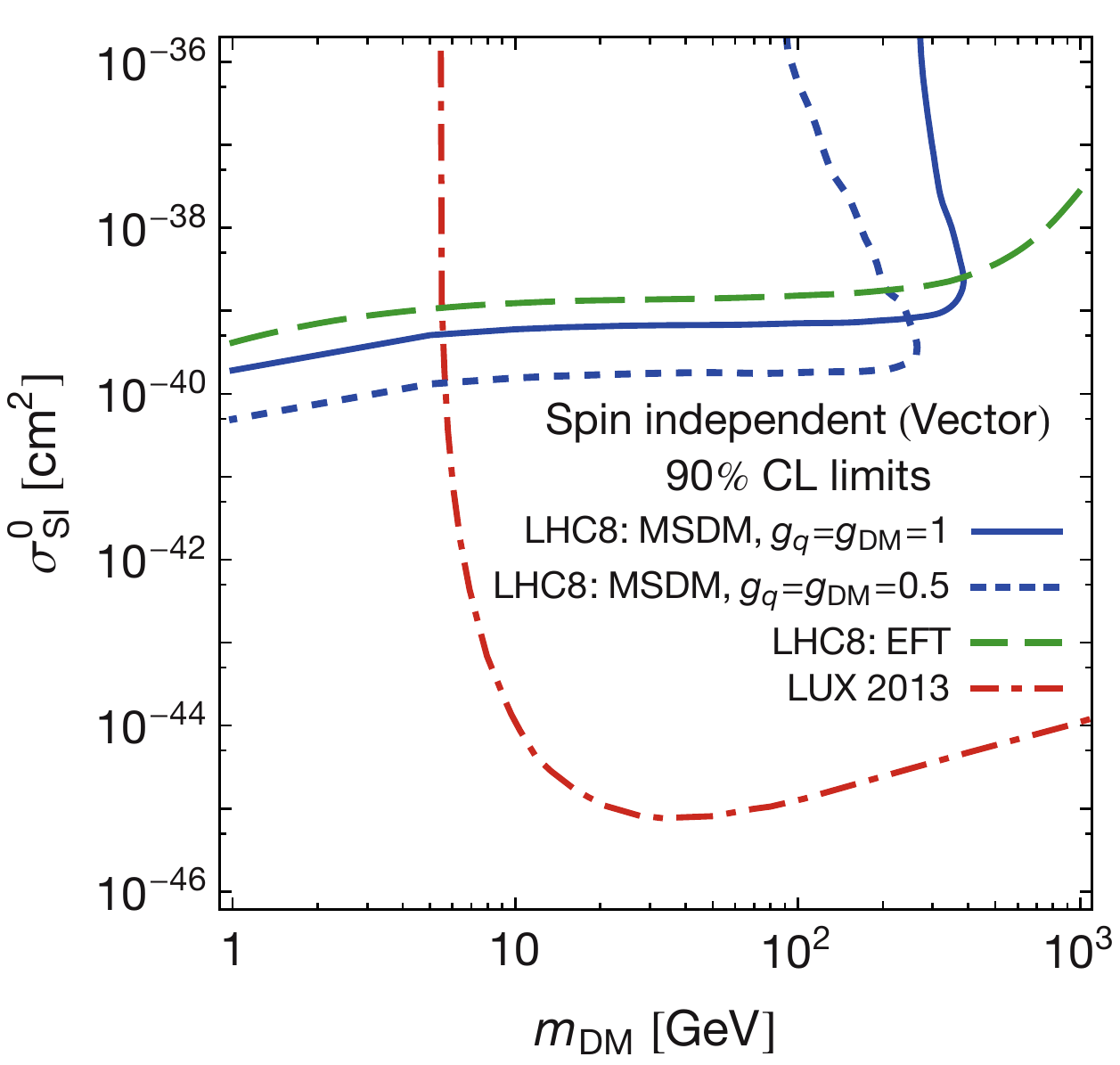}
\includegraphics[width=0.495\columnwidth]{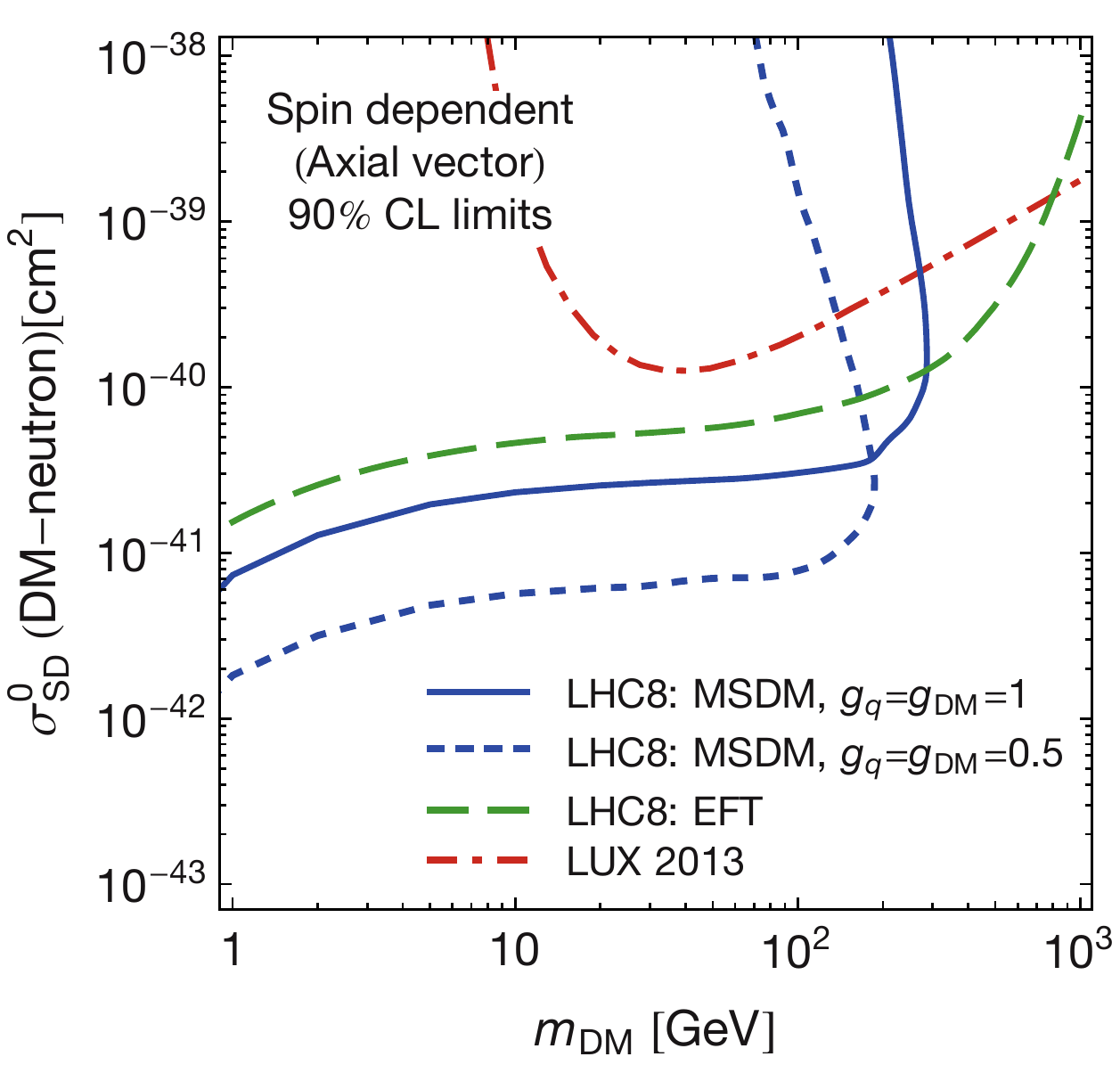}
\caption{A comparison between the 90\% CL mono-jet limit in our MSDM models (blue lines) and the EFT framework (green dashed) in the cross-section vs $\mDM$ plane used by the direct detection community. The left and right panels show the limits on the SI and SD cross-sections appropriate for vector and axial-vector mediators respectively. The red dot-dashed line shows the current LUX limit. The MSDM and EFT limits should agree in the domain where the EFT framework is valid. For these choices of parameters, the EFT limit underestimates the MSDM limits for $\mDM\lesssim300$~GeV and overestimates them for $\mDM\gtrsim300$~GeV. The EFT limit gives a misleading representation
of the relative sensitivity of mono-jet and direct detection searches.
} 
\label{fig:EFTcomparison2}
\end{figure}

For completeness, we also show in figure~\ref{fig:EFTcomparison2} a comparison of the MSDM and EFT limits in a format which may be more familiar. Here we map the MSDM limits for the cases $g_q=\gDM=1$ (solid blue line) and $g_q=\gDM=0.5$ (short dashed blue line) onto the usual cross-section vs DM mass plane used to present direct detection limits. The translation of the MSDM limits to the cross-section vs DM mass plane is performed by passing the mono-jet limits from figure~\ref{fig:Mdm-Mmed} through eqs.~\eqref{eq:SI2MSDM} and~\eqref{eq:SD2MSDM}. The left and right panels show the SI and SD cross-sections appropriate for vector and axial-vector mediators respectively. The red dot-dashed lines show the LUX limits and the long dashed green line shows the EFT limits.

We find that the EFT limit underestimates the MSDM collider limit by almost an order of magnitude for $\mDM\lesssim300$~GeV (for $g_q=\gDM=0.5$) and overestimates the MSDM limit for $\mDM\gtrsim300$~GeV. As was discussed in section~\ref{sec:current},  the MSDM mono-jet production cross-section is resonantly enhanced in the region $\mDM\lesssim300$~GeV. This enhancement is not accounted for in the EFT limits, which explains why the MSDM limit is more constraining than the EFT limit in this mass range.

The size of the enhancement in the 90\%~CL limit of scattering cross-sections $\sigma^0_{\rm{SI}}$ and $\sigma^0_{\rm{SD}}$ relative to the EFT limit can be estimated using the `rules of thumb' in Appendix~A of~\cite{Buchmueller:2013dya}. The MSDM and EFT scattering cross-sections are approximately related by $\sigma_{\rm{MSDM}} \approx \left(\Gamma_{\rm{med}}/ \mMed\right)  \sigma_{\rm{EFT}}$.  For the $g_q=\gDM=1$ case, the ratio of the mediator width to the mediator mass is about 1/2, while for the $g_q=\gDM=0.5$ case it is closer to 1/8. Accordingly we expect the MSDM direct detection limit to be lower than the EFT limit by a factor~2 and~8 respectively in the regime where the resonant enhancement occurs i.e.~at low to moderate DM masses. We see from figure~\ref{fig:EFTcomparison2} that this relation is a good approximation when $\mDM \lesssim 300$~GeV for both the vector and axial-vector cases.

The EFT limits shown in figures~\ref{fig:EFTcomparison} and~\ref{fig:EFTcomparison2} are the result of a naive application of the limits on $\Lambda$ from figure~\ref{fig:validation}. Of course, in reality the EFT approach is only expected to be valid in a limited region of parameter space and various criteria have been proposed to designate the region where the EFT description is not valid. Perhaps the most naive criterion is to assume that the EFT limits are valid only when $\Lambda>\mDM/(2\pi)$ (see e.g.~\cite{Goodman:2010yf}). This is a very weak criterion and does not restrict any of the EFT limits in figures~\ref{fig:EFTcomparison} and~\ref{fig:EFTcomparison2}. When applying the EFT limits to a more-complete model, a more reasonable minimum criterion for the validity of the EFT approach is to demand that $\mMed > 2\mDM$. We have included the line $\mMed =2\mDM$ in the upper left panel of figure~\ref{fig:EFTcomparison} to demonstrate how this restricts the domain of the EFT limit. In this case, only the dashed green line below the $\mMed =2\mDM$ line is valid. While we see that this criterion excludes the EFT limit which differs most from the underlying MSDM limit (at large $\mDM$), we find that the EFT limit in the valid region still fails to accurately reproduce the MSDM limit in this part of parameter space. For instance, at $\mDM=100$~GeV the EFT limit on $\mMed$ underestimates the MSDM limit by 200~GeV. This highlights a general problem that even in regions where the EFT limit is valid, the limits are not model independent because the EFT framework does not include the mediator width,  which strongly affects the mono-jet limits.

In summary, important physical effects are missed in the EFT framework because the full effect of the s-channel mediator propagator is ignored.
 For instance, the limits in a full model may be substantially stronger than those found from the EFT limit on $\Lambda$ because the production can be resonantly enhanced. Furthermore,  the LHC has sensitivity to the underlying coupling structure because the production cross-section is sensitive to the mediator width (which breaks the degeneracy between~$\gq$ and~$\gDM$), but this is ignored because of the oversimplification of the EFT framework. The direct comparison of the MSDM and EFT limits highlights again how the EFT limits, when applied naively, lead to misleading conclusions about the real sensitivity of collider searches.  Thus, the EFT framework is unsuitable for quantifying the true complementarity of collider and direct detection searches.

\section{Conclusions}
\label{sec:conc}

In many previous studies, the effective field theory (EFT) framework has been utilised to interpret and characterise studies of dark matter (DM) production at the LHC. The EFT framework is very powerful in its domain of validity. Unfortunately, as we discussed in our previous paper~\cite{Buchmueller:2013dya}, the limits from the EFT framework for collider searches apply only to a limited class of theories in which the mediator mass is very heavy and the couplings are very large. In particular, in the region where the EFT is valid, the mediator width is often larger than the mass of the mediator ($\Gamma_{\rm{med}}>\mMed$), meaning that a particle-like interpretation of the mediator is difficult (in the context of a single mediator).  

In this paper we propose a Minimal Simplified Dark Matter (MSDM) framework, which is a more robust and accurate approach for interpreting and characterising collider searches of dark matter. In section~\ref{sec:SMS} we introduce MSDM models for vector and axial-vector mediators. In its most minimal variant our models are characterised by four free parameters: $\mDM,\, \mMed,\, \gDM$ and~$\gq$, which are the DM and mediator masses, and the mediator couplings to DM and quarks respectively. These parameters are sufficient to fully characterise DM production at colliders and scattering at direct detection experiments (see figure~\ref{fig:OP}). The advantage of the MSDM models is that the full event kinematics are captured and the dependence on all couplings and masses can be systematically studied.

After validating our implementation of the CMS mono-jet and LUX direct detection searches (see figures~\ref{fig:validation} to~\ref{fig:DDvalidation}), we map out the four-dimensional parameter space of our MSDM models by showing projections in two parameters. For vector mediators, we find that generally the LUX limits are much more constraining than the mono-jet limits (see left panels in figures~\ref{fig:Mdm-Mmed} to~\ref{fig:couplings}). The only exception is when $\mDM\lesssim5$~GeV, where direct detection experiments lose sensitivity. In this DM mass range the LHC limits are more constraining (see figure~\ref{fig:low-mass}). In contrast, the LHC and LUX limits on axial-vector mediators generally show full complementarity, probing orthogonal directions in the parameter space (see right panels in figures~\ref{fig:Mdm-Mmed} to~\ref{fig:couplings}). For instance, the mono-jet search probes larger values of $\mMed$ while direct detection searches probe larger values of $\mDM$.

We also provide estimates for the projected limits from the LHC for 14~TeV operation after $30~\ifb$, $300~\ifb$ and $3000~\ifb$, and from LZ after a 10 ton year exposure (see figure~\ref{fig:proj:Mdm-couplings}). It is interesting to note that the mono-jet reach in the axial-vector case approaches the neutrino noise, which, with current technology and calculations, is considered an irreducible background for direct detection experiments.  It therefore seems critical to combine both search approaches in order to have the best possible coverage for discovery in the future. 
 
 We further explore the validity of the EFT framework by comparing the limits from our MSDM models with the EFT limits (see figures~\ref{fig:EFTcomparison} and~\ref{fig:EFTcomparison2}). The EFT limits fail to give a good approximation to the  MSDM limits for both vector and axial-vector mediators over almost all of the parameter values considered. Confirming the results in~\cite{Buchmueller:2013dya}, we find that the EFT limits give a good representation of the MSDM limits only in the case of heavy mediator mass and large couplings. The EFT limits may also easily lead to misleading conclusions regarding the complementarity of collider and direct detection searches.

The MSDM framework is easily extendable to include scalar and pseudo-scalar interactions, Majorana fermion and scalar DM particles, as well as limits from indirect DM searches and additional collider searches, such as di-jet and $\rm{jets}+\MET$ searches. With these additions it should be possible to perform a global fit to a MSDM model, similar to the approaches already performed for supersymmetry (see for example~\cite{AbdusSalam:2009tr,Allanach:2011ya,Buchmueller:2013rsa}), to allow a more quantitative definition of the complementarity of collider and direct detection searches. This will become especially important in case of a discovery of a dark matter signal either at the LHC or at a direct detection experiment or ideally, at both.

\section*{Acknowledgements}

We thank Celine B\oe hm, Felix Kahlhoefer and Valya Khoze for productive discussions, and Emanuele Re for his extensive help with POWHEG throughout this project. Additionally, CM thanks Julien Billard, Enectali Figueroa-Feliciano and participants of the NORDITA `What is the dark matter?' program for helpful discussions regarding the limiting impact of coherent neutrino scattering on direct detection discovery. We also acknowledge the useful discussions at the Dark Matter brainstorming workshop at Imperial College London, and especially are grateful for input to our paper from Albert De Roeck and Greg Landsberg.

\bibliography{ref}
\bibliographystyle{JHEP}

\end{document}